\documentclass[aps,pre,reprint,amsmath,amssymb,showpacs,showkeys,superscriptaddress,pdftex]{revtex4-1}

\usepackage{graphicx}
\usepackage{hyperref}
\usepackage{xcolor}
\usepackage{array}
\usepackage{url}

\newcommand{\Eref}[1]{Eq.~\eqref{#1}}
\newcommand{\Fref}[1]{Fig.~\ref{#1}}
\newcommand{\Tref}[1]{Tab.~\ref{#1}}
\newcommand{\etal}{~{\it et al.}}
\newcommand{\nc}{n^\textrm{c}}
\newcommand{\nt}{n^\textrm{t}}
\newcommand{\nNW}{n^\textrm{NW}}
\newcommand{\ncs}{n^\textrm{cs}}
\newcommand{\el}{\ell}
\newcommand{\ec}{c}
\newcommand{\Wa}[3]{W#1/#2-#3}
\renewcommand{\S}{S_{2}}
\newcommand{\C}{C_{2}}

\newcommand\vek[1]{\mathbf{#1}}

\newcommand{\vfrac}{\phi}

\newcommand\de[1]{\,{\mathrm d}#1}
\newcommand{\Se}{\widehat{S}_2}

\newcommand{\iqa}{{IQA}}
\newcommand{\iqam}{{IQAM}}
\newcommand{\cshd}{{CSHD}}
\newcommand{\erro}{E(i,j)}
\newcommand{\overlap}{\omega}
\newcommand{\boundingBox}{b}

\newcommand{\puc}{PUC}
\newcommand{\set}[1]{{\mathbb #1}}
\newcommand{\setN}{\set{N}}

\newcommand{\measure}[1]{|#1|}\newcommand{\SII}{\ensuremath{S_2}}
\newcommand{\CII}{\ensuremath{C_2}}

\newcommand{\Ephi}{E_{\phi}}
\newcommand{\EQ}{E_{Q}}
\newcommand{\EC}{E_{C}}
\newcommand{\rsamples}{\mbox{r-samples}}
\newcommand{\rsample}{\mbox{r-sample}}
\newcommand{\rsampleHeight}{h}

\renewcommand{\SS}{\mathrm{RS}}

\newcommand{\TS}{\mathrm{TS}}
\newcommand{\hdisks}{{h-discs}}
\newcommand{\pdisks}{{s-discs}}
\newcommand{\sandstone}{{sandstone}}
\newcommand{\alporas}{Alporas\textsuperscript{\textregistered}}

\DeclareGraphicsExtensions{.png,.pdf,.eps,.jpg}

\begin{document}

\title{Aperiodic Compression and Reconstruction of Real World Material Systems Based on Wang Tiles}

\author{Martin Do\v{s}k\'{a}\v{r}}
\email{martin.doskar@fsv.cvut.cz}
\affiliation{%
  Department of Mechanics,
  Faculty of Civil Engineering,
  Czech Technical University in Prague,
  Th\'{a}kurova 7,
  166 29 Praha,
  Czech Republic
}
\author{Jan Nov\'{a}k}
\email{novakj@cml.fsv.cvut.cz}
\affiliation{%
  Department of Mechanics,
  Faculty of Civil Engineering,
  Czech Technical University in Prague,
  Th\'{a}kurova 7,
  166 29 Praha,
  Czech Republic
}
\affiliation{%
  Institute of Structural Mechanics,
  Faculty of Civil Engineering,
  Brno University of Technology,
  Veve\v{r}\'{i} 95/331,
  602 00 Brno,
  Czech Republic
}
\author{Jan Zeman}
\email{zemanj@cml.fsv.cvut.cz}
\affiliation{%
  Department of Mechanics,
  Faculty of Civil Engineering,
  Czech Technical University in Prague,
  Th\'{a}kurova 7,
  166 29 Praha,
  Czech Republic
}

\begin{abstract}
The paper presents a concept/technique to compress and synthesize complex material morphologies that is based on Wang tilings. Specifically, a microstructure is stored in a set of Wang tiles and its reconstruction is performed by means of a stochastic tiling algorithm. A substantial part of the study is devoted to the setup of optimal parameters of the automatic tile design by means of parametric studies with statistical descriptors at heart. The performance of the method is demonstrated on four two-dimensional two-phase target systems, monodisperse media with hard and soft discs, \sandstone{}, and high porosity metallic foam.
\end{abstract}

%
%


\maketitle

\section{Introduction}\label{s:introduction}

Randomness and heterogeneity govern majority of real world processes. Indeed, materials which are considered as macroscopically homogeneous usually display heterogeneity across multiple scales, e.g. defects in crystalline or quasi-crystalline lattice, imperfect or entirely random particle packings, random distribution of pores or inclusions, etc.~\cite{Torquato:2002}. Regarding the incorporation of morphological information in multiscale simulations, it has progressed namely through a unit cell definition for heterogeneous materials exhibiting locally periodic arrangement of constituents~\cite{Michel1999109,Geers20102175}. Moreover, the unit cell approach has also penetrated into the framework of random microstructures, namely due to the lack of an appropriate alternative, e.q.~\cite{Povirk:1995:IMI,zeman2007random,Owhadi2003}. In this paper, we thus bring a generalization to the single unit cell concept that is based on Wang tiles and tailored especially for computer simulations of disordered material systems.

The concept of Wang tiles was introduced by Hao Wang in 1961 as a method to decide whether a certain class of logical statements can be proven by means of axioms of mathematical logic encoded in planar patterns~\cite{wang1961tiling,wang1965games}.
Afterwards, a focus was on the discovery of a finite set of tiles that could tile the infinite plane aperiodically in order to find counterexamples to Wang's decidability conjecture on statements mirrored in periodic sets\footnote{``\dots What appears to be a reasonable conjecture, which has resisted proof or disproof so far (as of 1961) is: 4.1.2 The fundamental conjecture: A finite set of plates (meaning tiles) is solvable (has at least one solution) if and only if there exists a cyclic rectangle of the plates; or, in other words, a finite set of plates is solvable if and only if it has at least one periodic solution.''~\cite{wang1961tiling}}.

The first result in this direction was obtained by Berger in 1966, who established a relation between aperiodic Wang tilings and the Turing-Davis halting problem and introduced the first finite aperiodic set consisting of 20,426 tiles~\cite{berger1966undecidability}, reduced to 104 later on~\cite{grunbaum1986tilings}. Further developments were brought by Amman~\cite{grunbaum1986tilings}, who originated the discovery of the set of 16 tiles, and by Culik and Kari~\cite{Culik:1996:ASW:245761.245814}, who scored 13. Kari and Culik also introduced an extension to three dimensions by means of Wang cubes~\cite{culik1995aperiodic}.

In addition to the strictly aperiodic constructions, tile sets that are not aperiodic themselves may allow for aperiodic tilings in a stochastic sense. This was first recognized by Cohen\etal{}~\cite{cohen2003wang}, who used Wang tiles to produce irregular patterns for purposes of computer graphics as an extension of Stam's seminal idea based on aperiodic Amman's set~\cite{Stam1997}. Conforming edge information of Wang tiles was also instrumental in syntheses of biological motifs. In particular, the tiles were exploited for the assembly of DNA double-crossover molecules with edges playing the role of peptide bonds according to the Watson-Crick molecular complementarity~\cite{winfree1998design}. In addition, the motifs based on DNA branched junctions were used together with tiles for self-assembly of aperiodic scaffolds in order to produce devices or specifically structured nanometre grids~\cite{Yan26092003}.

Wang tiles have also found use in Statistical Physics. To name a few, Amman's set was used for the modelling of nucleation of a metastable quasicrystalline phase~\cite{aristoff2011first} whose formation depends on the cooling rate of the alloy melt~\cite{shechtman1984metallic}. The model was based on the thermodynamics of Wang tilings studied earlier by Leuzzi\etal~\cite{leuzzi2000thermodynamics}.

As for the micromechanics of materials, the application of Wang tiles to the compression of geometry of particulate suspensions was proposed in~\cite{NovakPRE2012}. In this work, the tiles carry microstructural patterns designed to minimize a difference between spatial statistics of a target system and the reconstructed media. Preliminary results for tiles carrying patterns of mechanical fields were reported in~\cite{NovakMSMSE2012}.

In this paper, we further examine the potential of Wang tiles in compression of real world material systems. To this goal, tiles whose design rests on image fusion techniques~\cite{efros2001image,cohen2003wang} are investigated and combined with statistics arguments in order to determine optimal setting of design parameters. The methodology is demonstrated on microstructural patterns of the media with uniformly distributed equi-sized hard and soft discs, sandstone, and closed cell aluminium foam \alporas{}. The proximity of synthesized microstructures to reference specimens is quantified by means of the one- and two-point probability functions, and the two-point cluster function.

\section{Background}\label{s:background}

Techniques combined in the present paper originate from various fields. A brief overview of relevant basics is given first as their omission would make the paper difficult reading.

\subsection{Concept of stochastic Wang tilings}

Basic elements of the concept are Wang (i) tiles, (ii) tile sets, and (iii) tilings. Wang tile is a square, jigsaw-like, quatrominoe piece capable of carrying arbitrary microstructural patterns within its entire domain including edges~\cite{wang1965games}. It is not allowed to be rotated or reflected when placed into a tiling, so that the tiles with an identical sequence of edges mutually rotated by $k\pi/4$ where $k\in\setN$ are considered different.
\begin{figure}[ht!]
  \centering
  \begin{tabular}{ccc}
    \includegraphics[height=2.2cm]{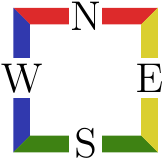} & 
    \includegraphics[height=2.2cm]{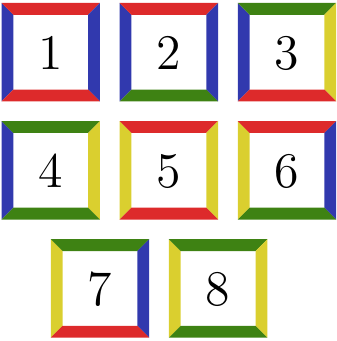} & 
    \includegraphics[height=2.2cm]{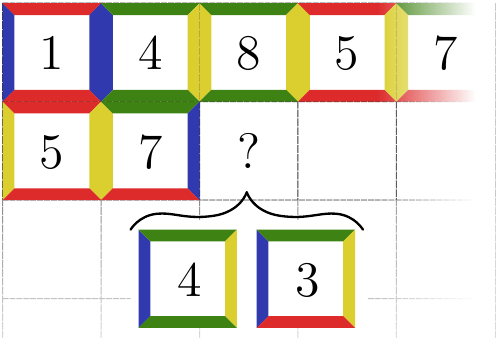} \\
    (a) & (b) & (c)
  \end{tabular}
  \caption{(Colour online) Wang tiling concept, a) tile with coloured edges, b) set \Wa{8}{2}{2}, i.e. $\nc_1=\nc_2=2$, $\nt=8$, $\nNW = 2$, $\ec_1\in\{\mathrm{green, red} \}$, $\ec_2\in\{\mathrm{blue, yellow} \}$  c) single step of \cshd{} algorithm}
  \label{fig:wang_tile}
\end{figure}

In applications, the edges are usually distinguished by colours~\cite{cohen2003wang}, alphabetical codes~\cite{NovakPRE2012}, or enumerated by integers~\cite{Culik:1996:ASW:245761.245814,aristoff2011first}, cf.~\Fref{fig:wang_tile}a. A collection of tiles that enables to cover up an open planar domain is called a tile set, \Fref{fig:wang_tile}b, here referred to as $\mathrm{W}\nt/\nc_1-\nc_2$, with cardinality $\nt$ and $\nc_i$ representing number of unique codes on horizontal ($i=1$) and vertical ($i=2$) edges, respectively~\cite{NovakPRE2012}. The magnitude of $\nt$ depends on the choice of $\nc_i$, so that it holds $\nt = \nNW\sqrt{\ncs}$, where $\nNW \in \set{M} = \{ 2, \dots ,\sqrt{\ncs} \}$. $\ncs$ stands for the number of all four-tuples given by the admissible permutations of edge codes $\ec_i$, or in other words, it refers to the cardinality of the so called complete stochastic Wang tile set. Having the above notion at hand, notice that the set \Wa{1}{1}{1} corresponds to a single periodic unit cell (\puc{}).

Reconstruction/synthesis of a piece of microstructure gives a tiling. It is a discrete mapping of tiles from the set onto the centres of a square planar lattice, where each tile conforms with its neighbours through coincident edge codes or is bounded from outside. In addition, we assume that there are no gaps in the tiling.

\begin{figure}[ht]
  \centering
  \begin{tabular}{cc}
    \includegraphics[width=0.45\columnwidth]{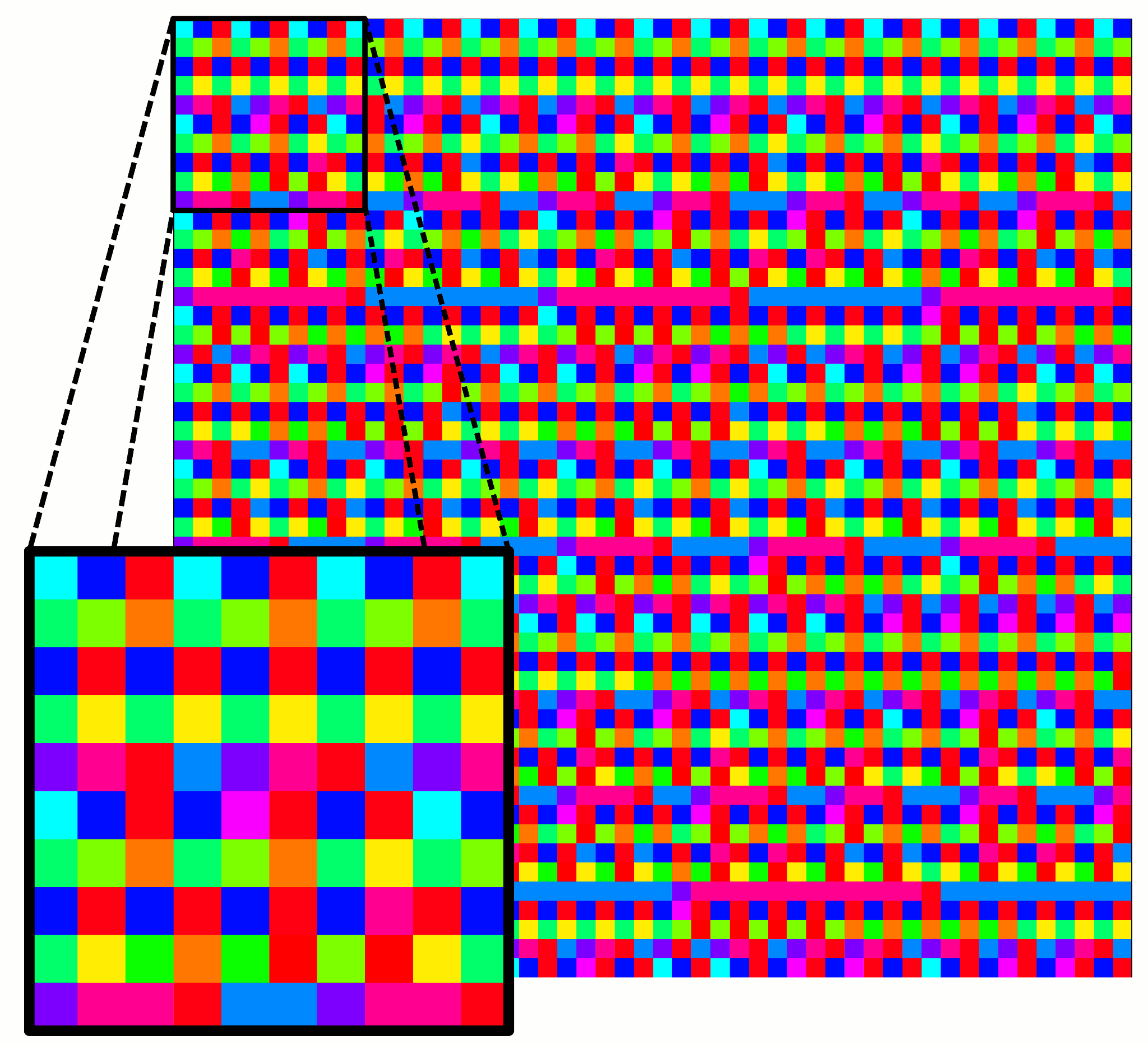}
    &\includegraphics[width=0.45\columnwidth]{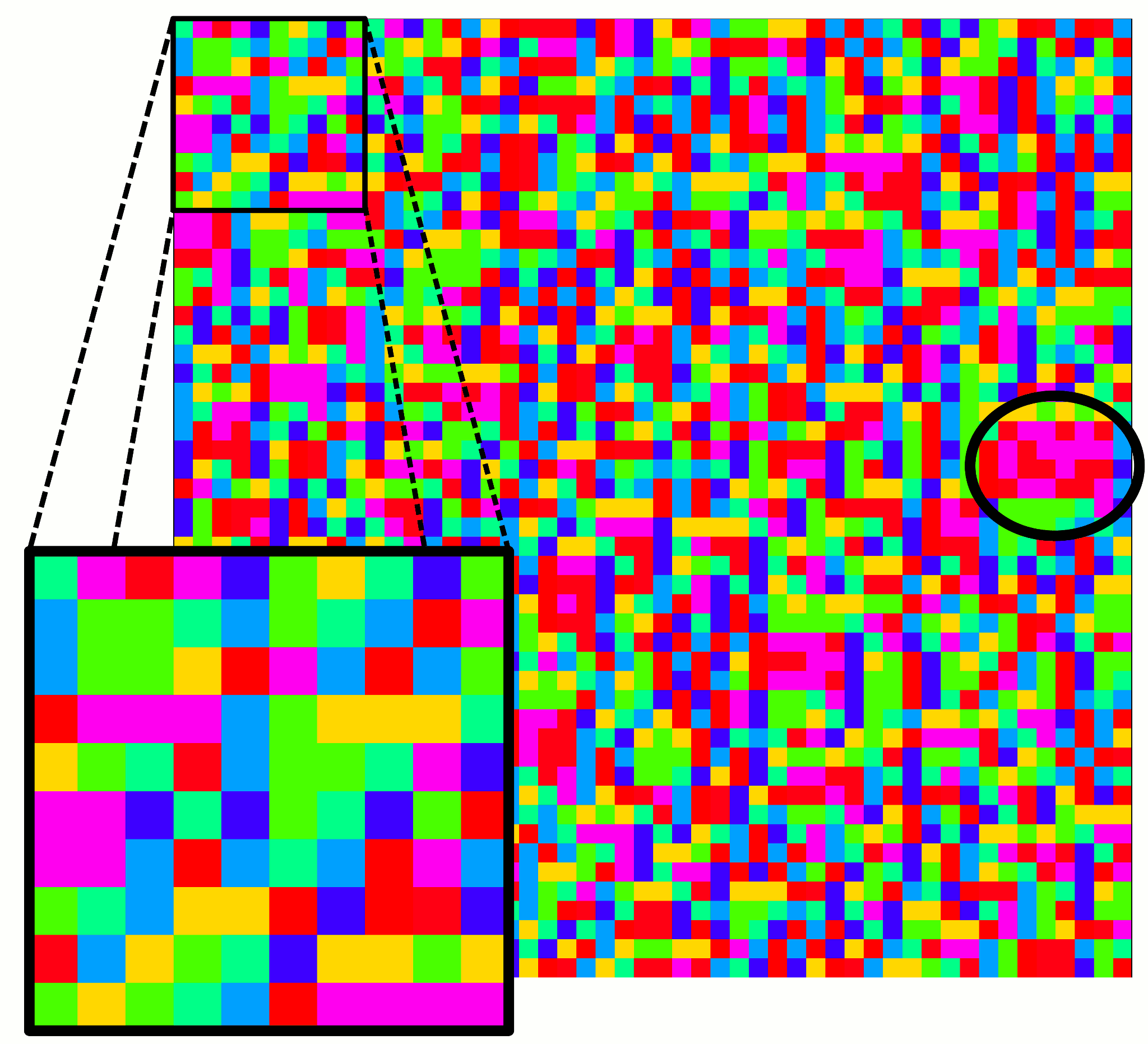} \\
    (a) &(b)
    \end{tabular}
  \caption{(Colour online) Wang tilings consisting of $50\times50$ tiles ($10\times10$ tiles in zoomed area) created by, a)~Kari-Culik 13-tile set and related cellular automaton, b) \Wa{8}{2}{2} and \cshd{} algorithm. Individual tiles are distinguished by solid colours}
  \label{fig:synthesis}
\end{figure}

When tiling the plane stochastically, the tiles are randomly selected from the set and successively placed one by one, either row-by-row or column-by-column, so that the edge codes of a newly placed tile must comply with those of its neighbours placed beforehand. Owing to the rectangular nature of Wang tiles, a pair of edges adjacent to north-western (NW) corner is controlled,~\Fref{fig:wang_tile}c. The index of the tile to be placed is selected randomly from the subset, which stores the tiles of identical NW edge code combinations. Recall, $\nNW$ must equal at least $2$, the minimal cardinality of the subset, in order to keep the procedure random. Aperiodicity of resulting tilings is guaranteed assuming the random number generator to never return a periodic sequence of numbers. We call the resulting algorithm \cshd{}, in honour of its authors~\cite{cohen2003wang}. In order to give an impression on the distinction between tilings made up of aperiodic and stochastic sets, respectively, compare results in~\Fref{fig:synthesis}a and \Fref{fig:synthesis}b. Clearly, the first method produces rather artificially looking patterns, while the latter leads to tilings with randomly distributed periodic clusters, e.g. the circled purple region in~\Fref{fig:synthesis}b. This phenomenon can be controlled by increasing $\nNW$, though at the expense of larger sets.

\subsection{Image quilting}

The proposed automatic design of tiles rests on the Image Quilting Algorithm (\iqa{}) due to Efros et al.~\cite{efros2001image} that allows for the fusion of raster images without severe visual defects. It seeks for a continuous path along which the desired pieces of microstructure are glued together, minimizing the sum of square differences of pixel values (1/0 for binary media) restrained to a certain overlap. Assume a pair of samples $A$ and $B$, both of the same height $\rsampleHeight$ and overlaying in a strip of the width $\overlap$. The local error in coincident pixels is defined as
\begin{multline}
  e(i, j) = [A(i, j) - B(i, j)]^2 \\ \mathrm{for} \quad (i, j) \in \{1,\dots,h\} \times \{1,\dots,\overlap\}
  \label{eq:pixel_error}
\end{multline}
It gives rise to the cumulative error
\begin{widetext}
\begin{equation}
  \erro = \left\{
	\begin{array}{ll}
		e(i, j) & i=1\\
		e(i, j) + \min\{E(i - 1, j - 1), E(i - 1, j), E(i - 1, j + 1)\} & i \in \{2,\dots,h\}
	\end{array}\right.
  \label{eq:cumulative_error}
\end{equation}
\end{widetext}
where the non-defined entries $E(i,0)$ and $E(i,\overlap+1)$ are excluded from consideration. The minimal cumulative error within the bottom row
\begin{equation}
	\EQ = E(h,Q(h)) = \min\{E(h,j), j \in \{1,\dots,\overlap\} \}
	\label{eq:overall_cumulative_error}
\end{equation}
thus characterizes defects caused by the image fusion and determines the horizontal coordinate of sought quilting path $Q(h)$ in the bottom row. The remaining path coordinates $Q(i)$, $i = 1,\dots,h-1$,  are found recursively by decrementing $i$
\begin{widetext}
\begin{equation}
	E(i,Q(i)) = \min\{ E(i,Q(i-1)-1), E(i,Q(i-1)), E(i,Q(i-1)+1) \}
\end{equation}
\end{widetext}

The simplicity of the algorithm is redeemed by noticing that the path can propagate only straight ahead or diagonally upwards. Moreover, it does not distinguish whether the path runs through inclusions or the matrix phase, yielding the inclusion shapes to deteriorate. This disadvantage, pronounced namely for binary media, can be reduced by recasting~\Eref{eq:pixel_error} as
\begin{equation}
  e(i, j) = \left\{ 
  \begin{array}{l}
    0,\quad \mathrm{for}\quad A(i, j) \wedge B(i, j)\, \textrm{in matrix}\,,\\
    1,\quad \mathrm{otherwise}
  \end{array}\right.
  \label{eq:error_modif_quilting}
\end{equation}
which leads to a modified algorithm referred to as \iqam{} in the sequel.

\subsection{Statistical quantification of microstructure}

Assuming a statistically homogeneous quasi-ergodic binary composite~\cite{sejnoha2013micromechanics}, we can best quantify the microstructure morphology by means of the $n$-point probability function $S_n$, which gives the probability of locating $n$ points $\vek{x}_1,\dots,\vek{x}_n$ in a given phase.  This phase corresponds to \emph{inclusions} in what follows. In particular, it is understood as the white phase for hard discs, soft discs and sandstone, and black phase for \alporas, see ahead \Fref{fig:target_systems}. The second phase is referred to as a \emph{matrix}.

Since we are primarily concerned with spatial correlations of multiples of the tile edge length $\el$, induced by the proposed compression framework, it is sufficient to limit the exposition to the two point probability $\S(\vek{x}_1,\vek{x}_2) = \S(\vek{x}_2-\vek{x}_1)$, which can be evaluated effectively in the Fourier space~\cite{Torquato:2002}.

Short range defects arising from the quilting technique will be investigated by the two-point cluster function $\C(\vek{x}_1,\vek{x}_2) = \C(\vek{x}_2-\vek{x}_1)$~\cite{jiao2009superior}, which can be understood as a special case of $\S$ function as it gives the probability of finding a pair of points $\vek{x}_1$ and $\vek{x}_2$ not only in the same phase, but also in the same cluster. The cluster is understood as the part of a phase where the two points $\vek{x}_1$ and $\vek{x}_2$ can be reached through a continuous path~\cite{torquato1988cluster}.
Thus, in addition to the information on distribution of inclusions in the matrix phase given by $\S$, $\C$ provides us with a short-range order description of inclusion shapes. A few limit cases can be distinguished. For distances $\measure{\vek{x}_2-\vek{x}_1}> b_{\mathrm{max}}$, where $b_{\mathrm{max}}$ refers to the maximum dimension of the largest inclusion, we have \mbox{$\C=0$}. The limit case $\measure{\vek{x}_2-\vek{x}_1}= 0$ yields $\C=\S=S_1= \vfrac$, where $\vfrac$ symbolizes the volume fraction (the one-point probability function).

\section{Automatic design of tiles}\label{s:methodology}
In order to arrive at microstructure compressions consistent from the statistical viewpoint, the techniques introduced above are combined. Contrary to our previous study based on an optimization approach~\cite{NovakPRE2012}, the automatic design of tiles is proposed as it suits better complex morphologies under consideration and is computationally more efficient.

\begin{figure}[ht!]
  \centering
  \setlength{\tabcolsep}{0.00cm}
  \begin{tabular}{>{\centering\arraybackslash}p{0.220\columnwidth}>{\centering\arraybackslash}p{0.420\columnwidth}>{\centering\arraybackslash}p{0.240\columnwidth}}
    \multicolumn{3}{c}{\includegraphics[width=0.98\columnwidth]{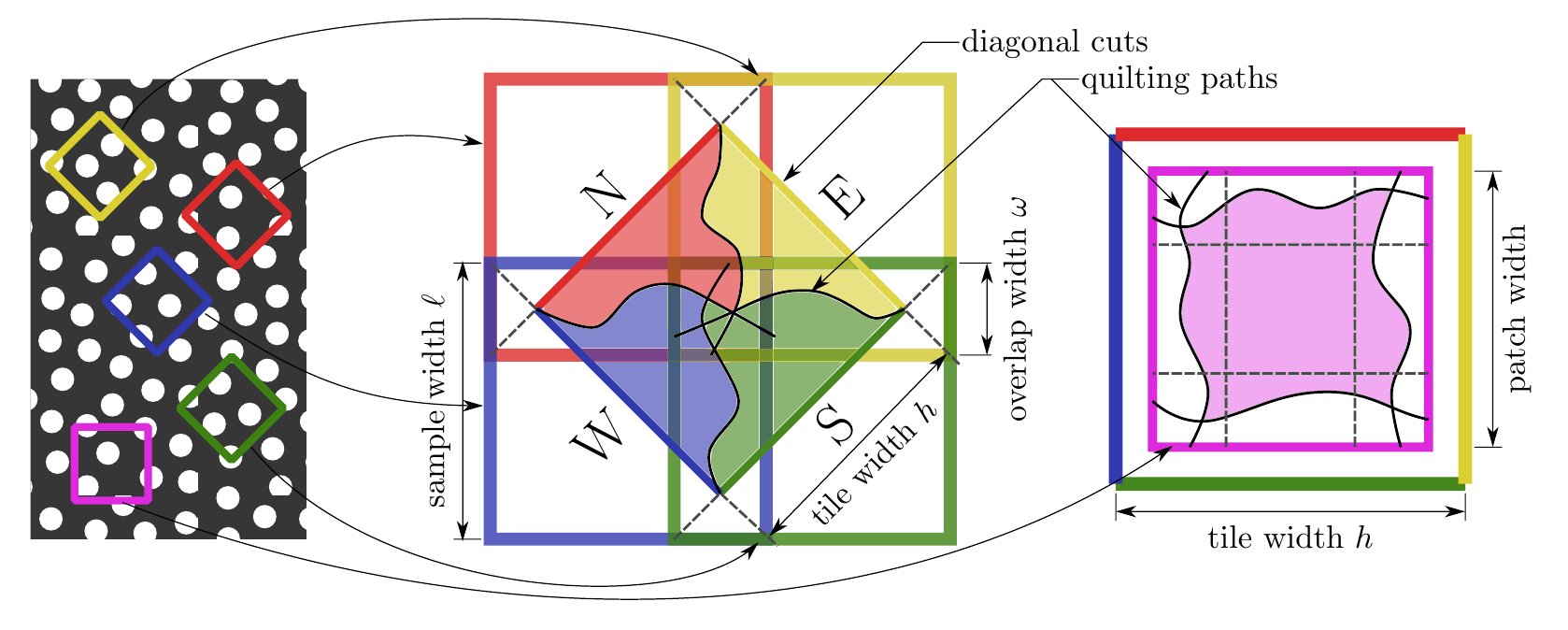}}\\
    (a) &(b) &(c)
  \end{tabular}
  \caption{(Colour online) (a, b) Illustration of Automatic tile design due to Cohen\etal{}~\cite{cohen2003wang}, and (c) proposed patch enrichment.}
  \label{fig:tiledesign}
\end{figure}

Following~\cite{efros2001image}, an automatically designed tile arises as a diamond shape cut out from the aggregate of four overlapping square reference samples, here called \rsamples{}, that are placed accordingly to the edge codes of the tile to be produced, \Fref{fig:tiledesign}. The four \rsamples{} are fused within an overlap $\overlap$ by means of the quilting algorithm and the resulting tile rotated by $\pi/4$. Its edge length yields from the \rsamples{} dimension $h$ and overlap width $\overlap$ as 
\begin{equation}
  \ell = \lceil\sqrt{2}(h-\overlap)\rceil
  \label{eq:edgeLength}
\end{equation}
where $\lceil\cdot\rceil$ denotes the round up operation to the nearest integer. 

The quilting paths among individual \rsamples{} always propagate from the tile corners towards the centre. The continuity of the tiling microstructure across the edges is ensured by the facts that the cut is taken diagonally across the r-sample, and that the same r-sample is used for all edges sharing the same code.

It has been shown in previous works that a specific morphology design technique may have an impact on the accuracy of the representation of long range orientation orders~\cite{NovakPRE2012,doskarUGT2012,NovakMSMSE2012}. Induced artefacts related to repeating tile edges or interiors may dominate spatial features of synthesized microstructures~\cite{NovakPRE2012}. Taking this into consideration, the automatic design procedure based on the fusion of the four \rsamples{} leads to compressions that strongly emphasize tile edges to the interiors. This is obvious from~\Fref{fig:tiledesign}b, noticing that almost the entire tile quarter is related to the edge information. A possible remedy proposed here is to replace a piece of the microstructure around the centre of automatically designed tiles by a square patch of the reference microstructure taken independently of the \rsamples{} and quilted around its perimeter as drawn in~\Fref{fig:tiledesign}c.

\subsection{Optimal overlap and best quilting performer}
\begin{figure*}[ht!]
  \centering      
    \includegraphics[width=\textwidth]{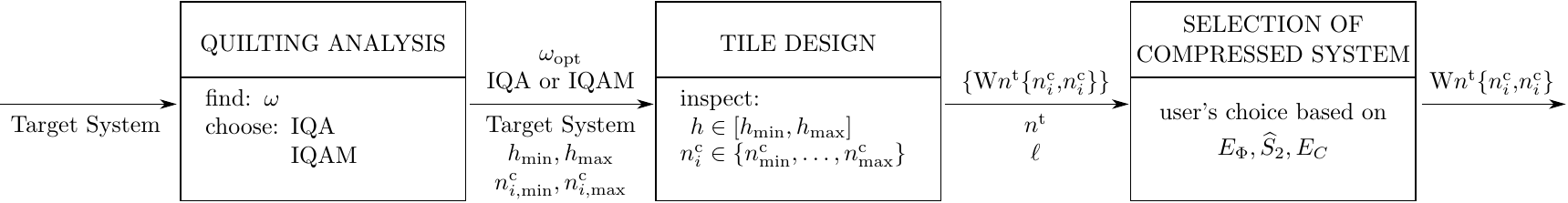}\\
    \caption{Sensitivity study flowchart to be passed through automatic tile design procedure. User defined input parameters are placed underneath arrow lines among individual cells. Outputs entering consecutive steps are written above them.}
  \label{fig:param_flowchart}
\end{figure*}
Four different target microstructures were considered in order to perform a sensitivity study on parameter values $\overlap, \rsampleHeight$, and $\nc_i$ with respect to the choice of the quilting algorithm \iqa{} or \iqam{},~\Fref{fig:param_flowchart}.

\begin{figure}[ht!]
  \centering      
  \begin{tabular}{cc}
    \includegraphics[width=0.48\columnwidth]{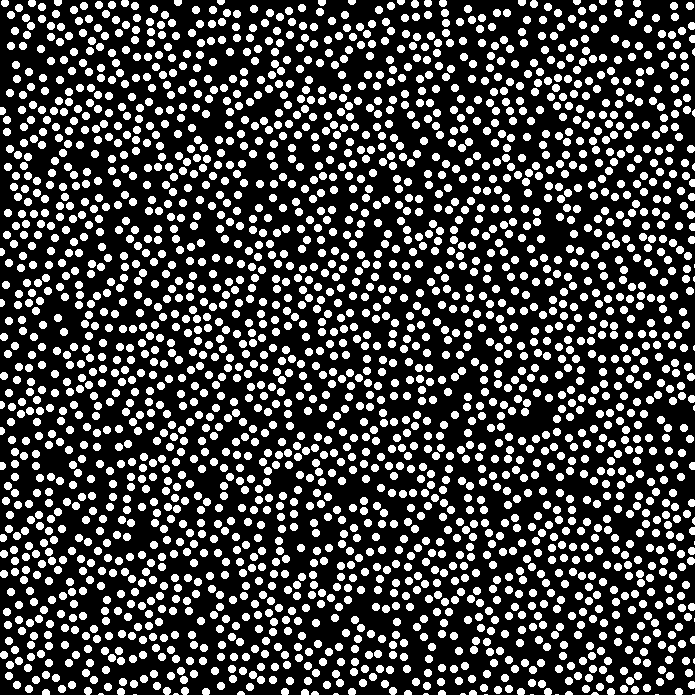}&
    \includegraphics[width=0.48\columnwidth]{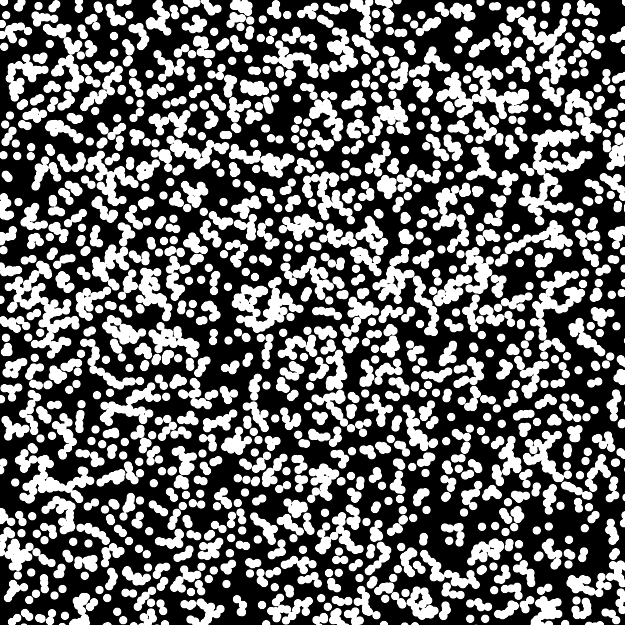}\\
    (a) & (b)\\
    \includegraphics[width=0.48\columnwidth]{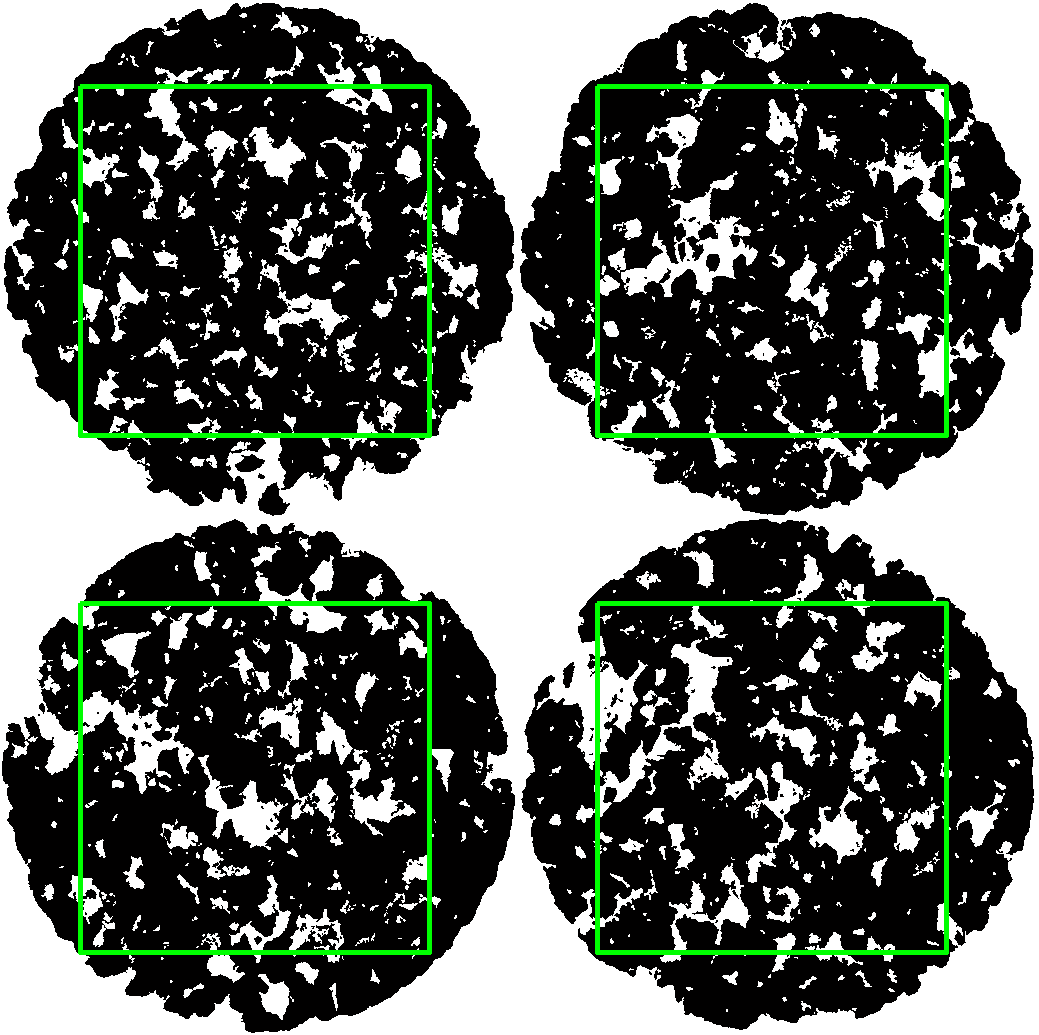}&
    \includegraphics[width=0.48\columnwidth]{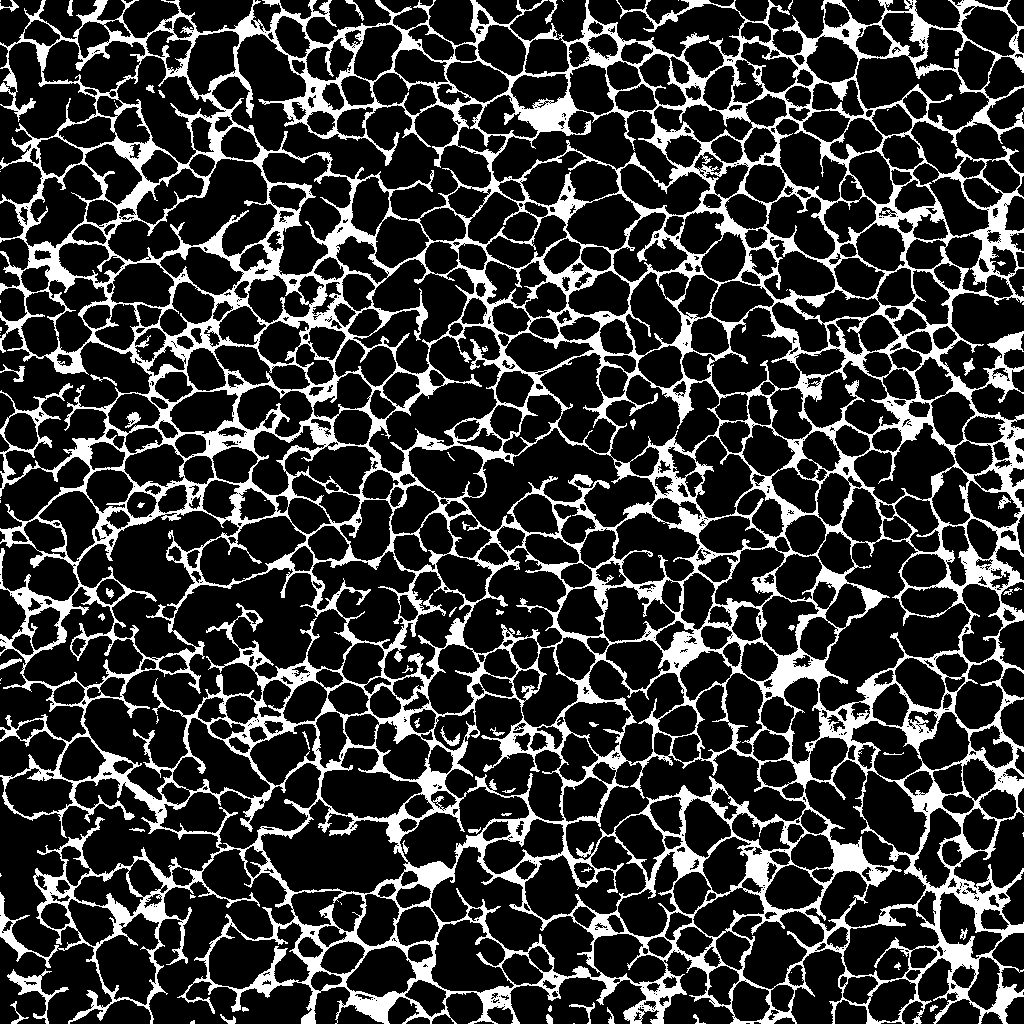}\\
    (c) & (d)
  \end{tabular}
  \caption{Target systems a) \hdisks, b) \pdisks, c) sandstone (Courtesy of Adrian Russell, UNSW, Sydney, Australia~\cite{sufian2013microstructural}), d) \alporas{} (Courtesy of Ji\v{r}\'{i} N\v{e}me\v{c}ek, CTU in Prague, Czech Republic~\cite{nvemevcek2013two})}
  \label{fig:target_systems}
\end{figure}

In particular, we have explored hard discs monodisperse (referred to as \hdisks{}),~\Fref{fig:target_systems}a, soft discs monodisperse (\pdisks{}),~\Fref{fig:target_systems}b, \sandstone{}, \Fref{fig:target_systems}c, and a large planar scan of \alporas{} foam,~\Fref{fig:target_systems}d. Notice that several cross-sections of sandstone CT data were used due to the insufficiency of microstructural information contained in a single one. 
                
The first aim was to find the optimal width of the overlap $\overlap$. For each material system, a pair of \rsamples{} of $h=300~\textrm{px}$ was chosen randomly from the reference microstructure. We varied the parameter $\overlap$ from $1$~pixel up to $10\times\boundingBox$~pixels, where $\boundingBox = \sqrt{\boundingBox_1^2 + \boundingBox_2^2}$, and $\boundingBox_i$ is the length of the mean inclusion bounding box in $i$th spatial direction. The minimum error path was sought after each increment and quantified according to \Eref{eq:cumulative_error}.
The whole process was repeated a hundred times to identify the sensitivity of the results to selection of different \rsample{} pairs, while keeping their height $h$ unchanged.
\begin{figure*}[htb!]
  \centering
  \setlength{\tabcolsep}{0.00cm}
  \begin{tabular}{ccc}
    \includegraphics[width=0.33\textwidth]{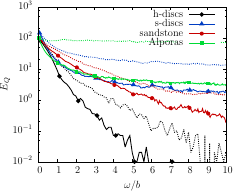}&
    \includegraphics[width=0.33\textwidth]{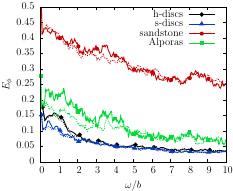}&
    \includegraphics[width=0.33\textwidth]{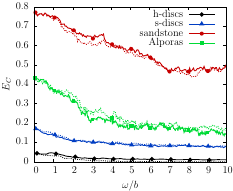}\\
    (a) & (b) &(c)\\
  \end{tabular}
  \caption{(Colour online) Optimal overlap width $\overlap$ with respect to \iqa{} (solid line), \iqam{} (dashed line), and various error measures. Plotted values represent means from $100$ realizations}
  \label{fig:overlap}
\end{figure*}

Quite surprisingly at first glance, \iqa{} outperforms \iqam{} in the sense of the norm defined by \Eref{eq:cumulative_error}. For example, it can be observed that for the target system consisting of \hdisks{},~\Fref{fig:overlap}a, \iqa{} returns zero-error quilting path for $\overlap/\boundingBox > 6$, whereas its modified version \iqam{} increases the value $\overlap/\boundingBox$ over $9.5$.  No zero cumulative error path $Q$ could be found for other material systems no matter the quilting algorithm we used. The errors exhibit similar asymptotic decay but the magnitude of the limit plateaus and oscillatory behaviour of the sandstone microstructure. Finding a quilting path as such is not the only problem related to the automatic design of tiles. Another difficulty arises from the random selection of \rsamples{} and from the fact that we take systematically into account only a portion of the microstructural information they contain. Moreover, the quilting algorithms bring additional very local defects into the morphology which influences the quality of the designed tiles. 

Therefore, we explored the quilting process also from the perspective of normalized deviation between phase volume fractions of the target and synthesized/reconstructed systems, $\vfrac^{\TS}$ and $\vfrac^{\SS}$, respectively,
\begin{equation}
  \Ephi= \frac{\measure{\vfrac^{\SS}-\vfrac^{\TS}}}{\vfrac^{\TS}}
  \label{eg:Ephi}
\end{equation}
and the relative error of the two-point cluster statistics defined as
\begin{equation}
  \EC = \frac{   \int_{(\overlap\times\rsampleHeight)} \measure{ {\C}^{\SS}(\vek{x}) - {\C}^{\TS}(\vek{x}) } \de{\vek{x}} } { \int_{(\overlap\times\rsampleHeight)} {\C}^{\TS}(\vek{x}) \de{\vek{x}} }
  \label{eq:EC}
\end{equation}

Both \iqa{} and \iqam{} demonstrate almost monotonic convergence in $\Ephi$ to a plateau value of about $0.05$ for all materials except \sandstone{} for which the error is still decreasing, however, we were not able to proceed with the analysis further due to the limited dimensions of input microstructures, \Fref{fig:overlap}b. Almost the same behaviour can be observed for $\EC$, \Fref{fig:overlap}c.\footnote{Degenerated inclusions at the boundary of overlap regions were excluded from calculations as they do not result from quilting algorithms.} It is also interesting to observe that the behaviour of \iqam{} in terms of spatial statistics is superior to the original version, namely for \alporas{} and \hdisks{}.

In summary, we can deduce that weakly packed dispersions are microstructural systems of the least complexity from the viewpoint of the automatic tile design, recall \Fref{fig:overlap}a. The performance of \iqa{} seems less powerful compared to the results provided by its modified version, however, both procedures are equivalent for remaining material systems and larger overlaps.
\begin{table}[htb!]
  \centering
  \begin{tabular}{l c c c c c}
    \hline\noalign{\smallskip}
    & & \hdisks{} & \pdisks{} & \sandstone{} & \alporas{}\\
    \hline \hline\noalign{\smallskip}
    $\overlap/\boundingBox$ &[-] & 5 & 5 & 6 & 6 \\
    $\overlap$ &[$\mathrm{px}$] & 40  & 75  & 120  & 180\\
    $\rsampleHeight$ &[$\mathrm{px}$] & 200  & 250  & 300  & 400\\
    $\el$ &[$\mathrm{px}$] & 227  & 107  & 255  & 312\\
    \hline
  \end{tabular}
  \caption{Optimal parameters of automatic design with respect to studied material systems}
  \label{tab:optimalvalues}
\end{table}

\subsection{Optimal tile edge length and cardinality of sets}

The optimal overlap $\overlap$, the second row of \Tref{tab:optimalvalues}, determines together with the dimension of \rsamples{} $\rsampleHeight$, length of tile edges $\el$, \Eref{eq:edgeLength}. In addition, the size of \rsamples{} depends on $\el$ and cardinality of sets given by $\nt$, dominant parameters in terms of the storage of required amount of microstructural information. 

To asses the optimal $\nt - \el$ combinations, various complete stochastic Wang tile sets were created. From each of those, subsets of $\nt$ tiles with minimum errors $\Ephi$ were chosen.

By analogy to the paragraphs above, $100$ realizations of $5\times 5$ tilings were synthesized by means of \cshd{} algorithm. The optimal setup of sought objectives, particularly $\el$ versus $\nt$, were assessed from the viewpoint of $\vfrac$, $\SII$ and $\CII$. In addition, we quantify the secondary extremes of $\SII$ by means of a set
\begin{equation}
  \Se = \left\{ \, \S( m \times \ell, n \times \ell),\, \forall (m,n) \in \set{Z}^2 \setminus (0,0) \, \right\}
  \label{eq:ES}
\end{equation}
\begin{figure*}[htb!]
  \centering
  \setlength{\tabcolsep}{0.00cm}
  \begin{tabular}{cccc}
    \includegraphics[width=0.245\textwidth]{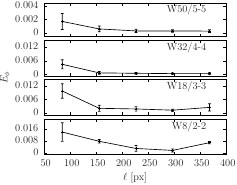}&
    \includegraphics[width=0.245\textwidth]{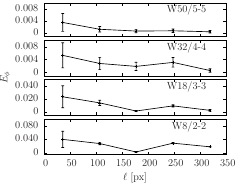}&
    \includegraphics[width=0.245\textwidth]{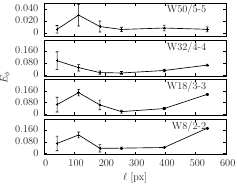}&
    \includegraphics[width=0.245\textwidth]{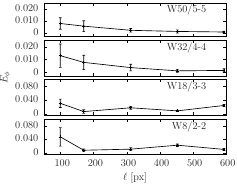}\\
    (a) & (b) & (c) & (d)    
  \end{tabular}
  \caption{Relative error of pore phase volume fraction versus tile edge length and cardinality of sets. The results are normalized against $\phi^{\TS}$, a) \hdisks{}, b) \pdisks{}, c) \sandstone{}, d) \alporas{}.}
  \label{fig:phi_vs_sample_size}
\end{figure*}

First, the behaviour of the relative error $\Ephi$ with respect to $\el$ is displayed for individual sets in \Fref{fig:phi_vs_sample_size}. From the graphs, it can be deduced that the scatter in volume fractions decreases for larger tiles and sets of higher cardinalities, independently of the material system. However, in some cases (\hdisks{} \Wa{8}{2}{2} and \Wa{18}{3}{3}, \sandstone{} all sets but \Wa{50}{5}{5}, and \alporas{} \Wa{18}{3}{3}) the error suddenly increases after the initial decay. In general, the reconstructions for \hdisks{}, \Fref{fig:phi_vs_sample_size}a, possess the least scatter from the target data, while the largest error is attributed to \sandstone{}, \Fref{fig:phi_vs_sample_size}c.
As for the higher order statistics, obtained results (not shown) proved the tile edge length $\el$ to have a negligible effect on reducing secondary extremes of $\S$, \Eref{eq:ES}, as well as no impact on the deviation between the two-point cluster functions of target and synthesized systems quantified by \Eref{eq:EC}, this time, with integrals over the domain of the tilings instead of $\overlap\times\rsampleHeight$.
\begin{figure*}[htb!]
  \centering
  \begin{tabular}{cc}
    \includegraphics[width=0.33\textwidth]{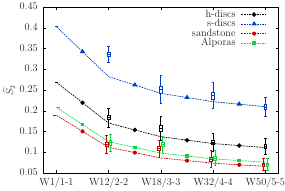} & \includegraphics[width=0.33\textwidth]{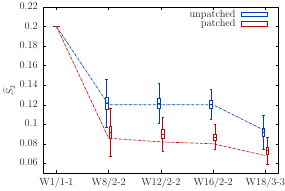}\\
    (a) & (b)
  \end{tabular}
  \caption{(Colour online) Reduction of secondary extremes of $\S$ with respect to number of edge codes, a) conventional tiles -- majority of microstructural information is carried by edges, b) comparison between conventional and patched tiles of \sandstone{} microstructure -- in patched tiles microstructural information is optimally distributed among edges and interiors. The boxes and whiskers, respectively, involve 50\% and 75\% of realizations proportionally distributed about the median. Continuous curves follow predictions given by~\Eref{eq:ES_prediction}.}
  \label{fig:sets_vs_S2}
\end{figure*}

On the other hand, the benefits of increasing the set cardinality are doubtless. In \Fref{fig:sets_vs_S2}a, we show the statistics of the secondary extremes $\Se$, \Eref{eq:ES}, by means of whiskerbar plots. The mean value of the set of secondary extremes $\Se$ seems to correspond well with the relation proposed in~\cite{NovakPRE2012}
\begin{equation}
  \Se^{\mathrm{p}} = \frac{\phi^t}{\nt} \left[ \vfrac + ( \nt - 1 ) \vfrac^2 \right] + \max_{i} \left\{ \frac{\phi^e}{\nc_i} \left[ \vfrac + ( \nc_i - 1 ) \vfrac^2 \right] \right\}
  \label{eq:ES_prediction}
\end{equation}
where $\phi^t$ and $\phi^e = 1 - \phi^t$ gives the portion of microstructural information attributed to tile interior and edges, respectively. We can clearly observe the proximity of simulated data to the estimate with $\phi^e=1$, which is attributed to the fact that the entire microstructural information in automatically designed tiles is associated with edges, contrary to the optimization based design~\cite{NovakPRE2012}. On top of that, observe a notably larger scatter (whiskers) in simulated data for individual tile set cardinalities compared to that discussed in~\cite{NovakPRE2012}. It stands to reason that the choice of tiles in explored sets preferred correct phase volume fractions to the uniform distribution of edges in synthesized tilings, which violates fundamental assumptions of \Eref{eq:ES_prediction} as justified in~\cite{NovakPRE2012}.
Therefore, the reconstructed microstructures are prone to repeat patterns specific to edges of higher frequencies, e.g. \Fref{fig:sinth_microstructures}c, thereby increasing $\Se$ in magnitude. A remedy consists in the use of patched tiles, compare Figs.~\ref{fig:sets_vs_S2}a,b and Figs.~\ref{fig:sinth_microstructures_patched} and~\ref{fig:sinth_microstructures}.

\section{Examples of compressed and synthesized microstructures}\label{s:examples}

\begin{figure}[htb!]
  \centering	
  \begin{tabular}{cc}
    \includegraphics[width=0.48\columnwidth]{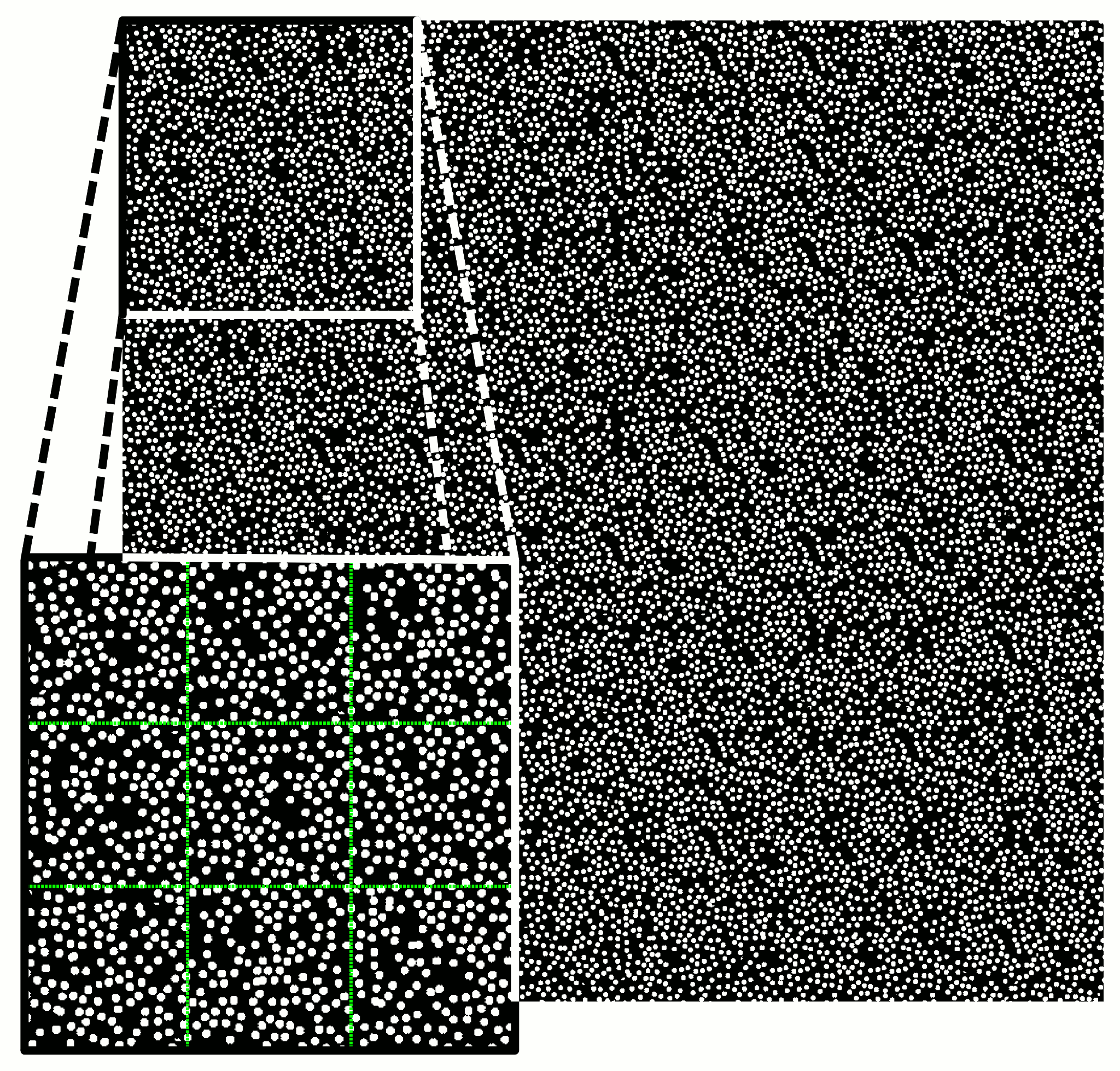}&
    \includegraphics[width=0.48\columnwidth]{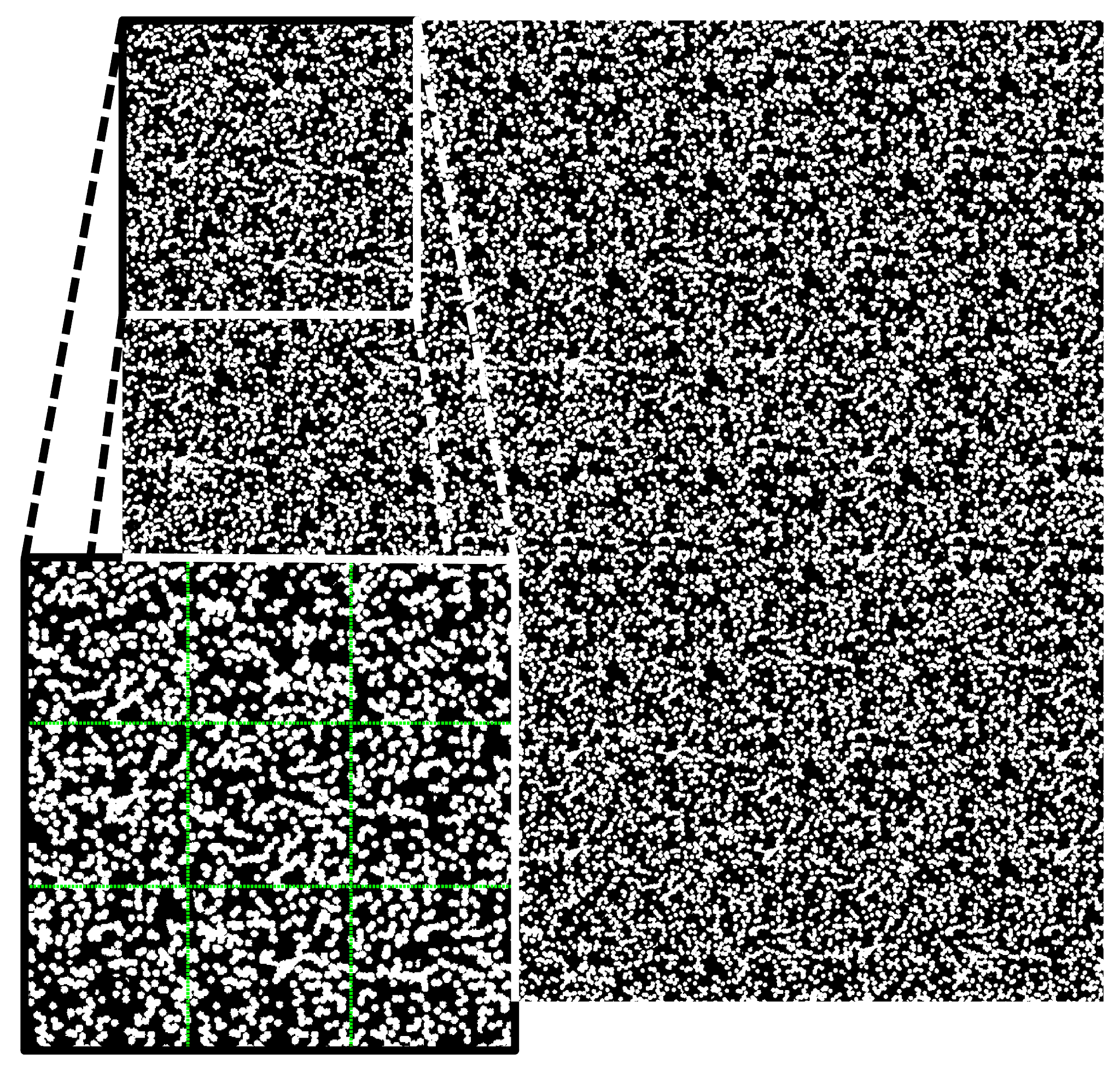}\\
    (a) & (b)\\
    \includegraphics[width=0.48\columnwidth]{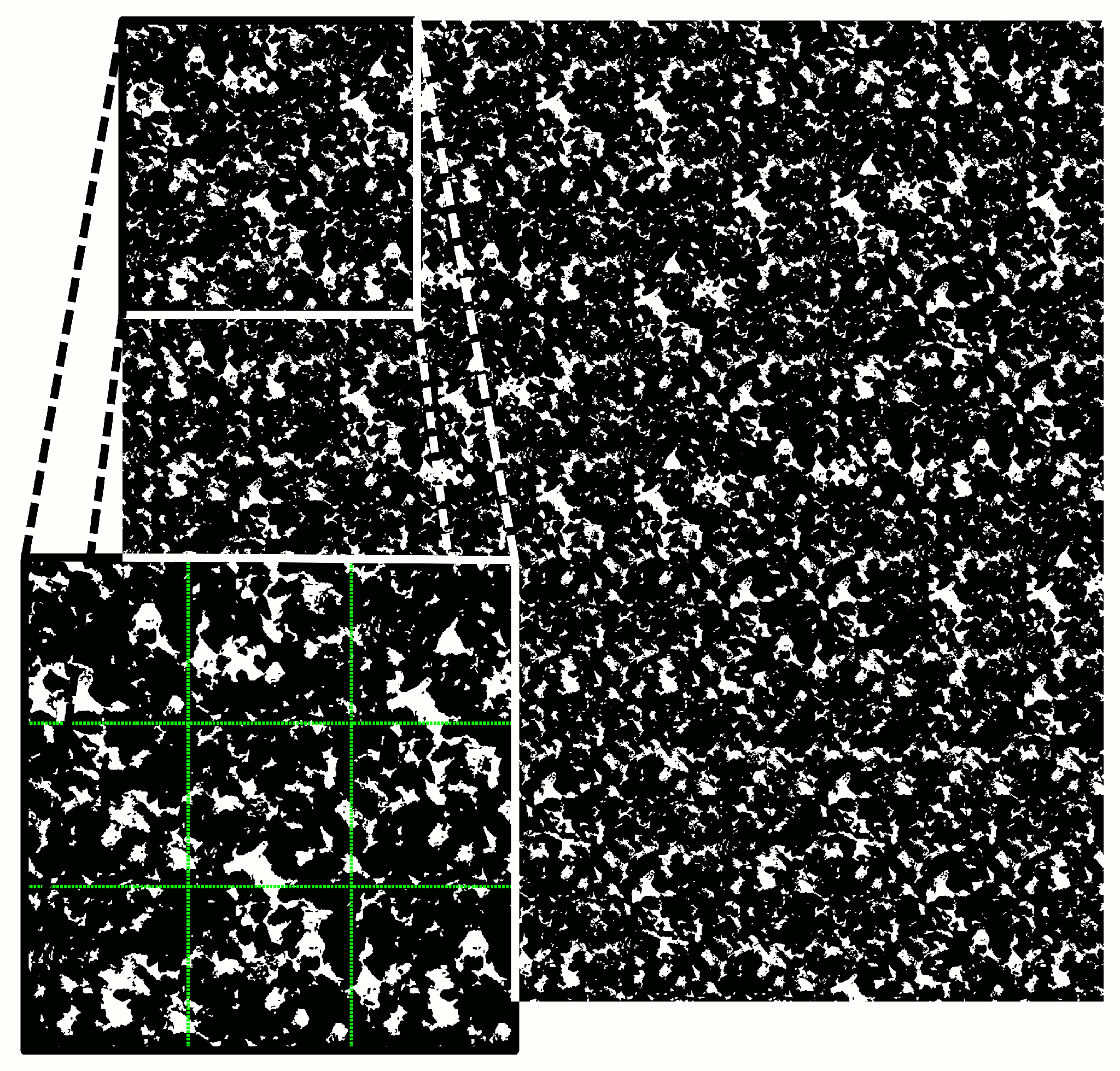}&
    \includegraphics[width=0.48\columnwidth]{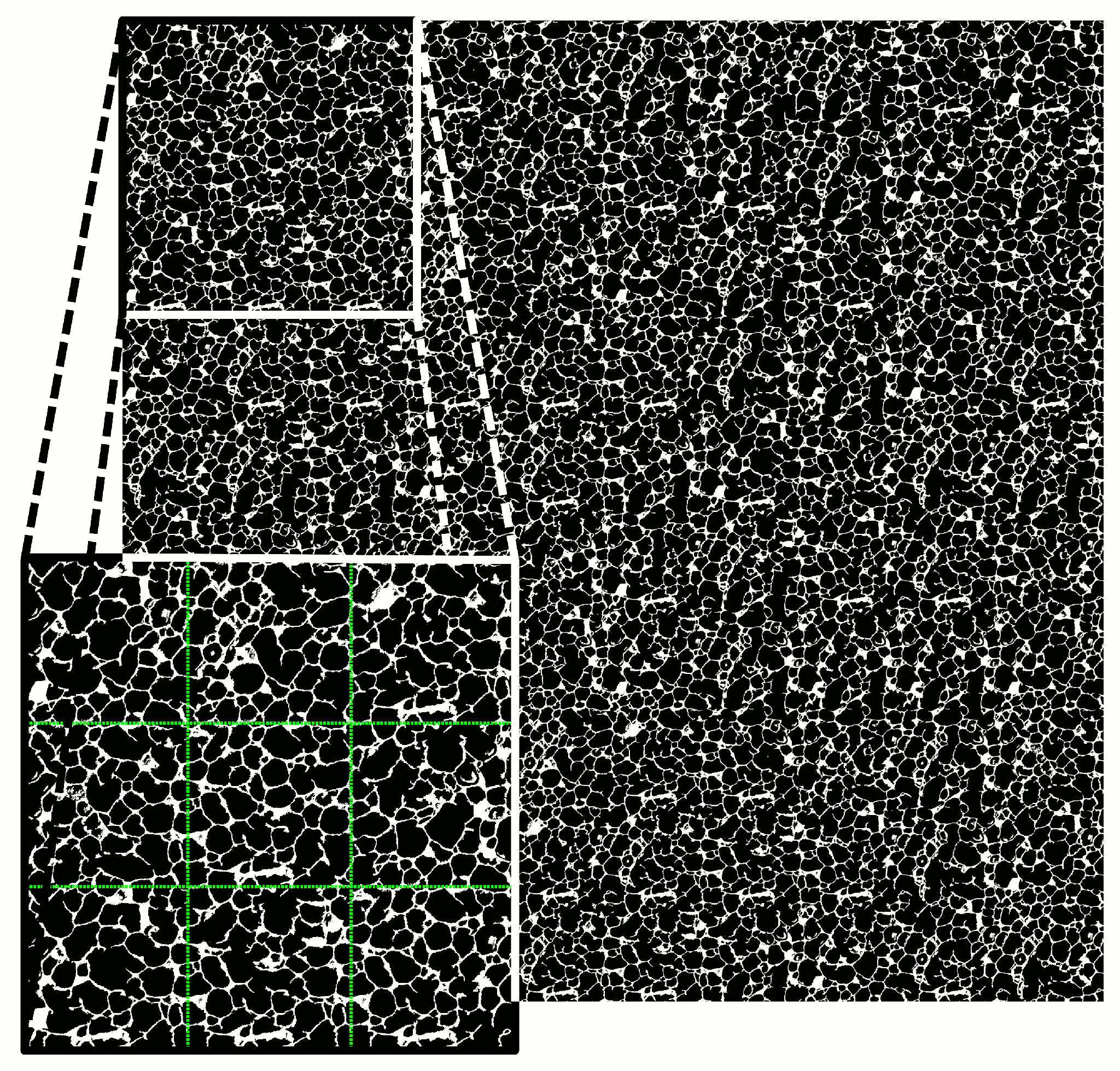}\\
    (c) & (d)
  \end{tabular}
  \caption{(Colour online) Examples of synthesized microstructure in $3\times3$ tiling, a) \hdisks{} (\Wa{50}{5}{5}), b) \pdisks{} (\Wa{50}{5}{5}), c) \sandstone{} (\Wa{50}{5}{5}), d) \alporas{} (\Wa{32}{4}{4})}
  \label{fig:sinth_microstructures}
\end{figure}

\begin{figure}[h!tb]
  \centering
  \begin{tabular}{cc}
    \includegraphics[width=0.48\columnwidth]{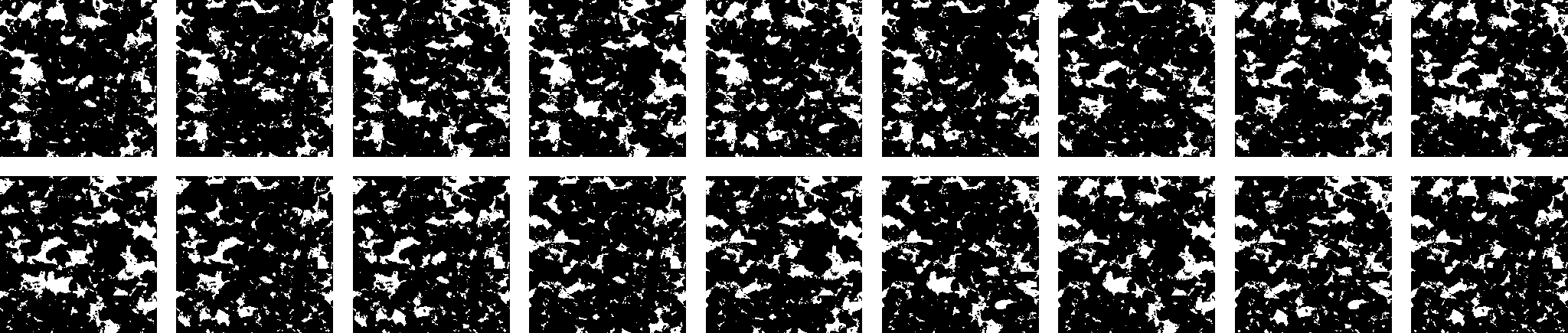}&
    \includegraphics[width=0.48\columnwidth]{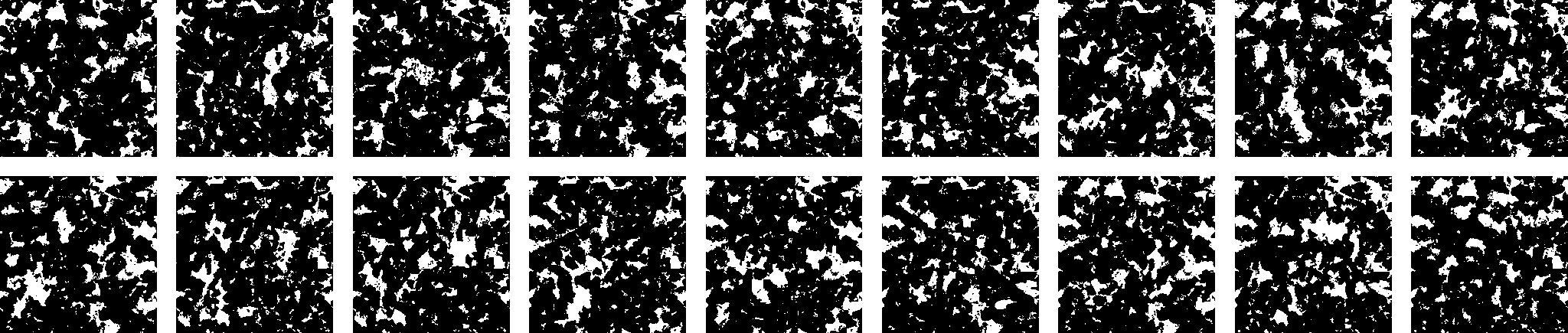}\\
    \includegraphics[width=0.48\columnwidth]{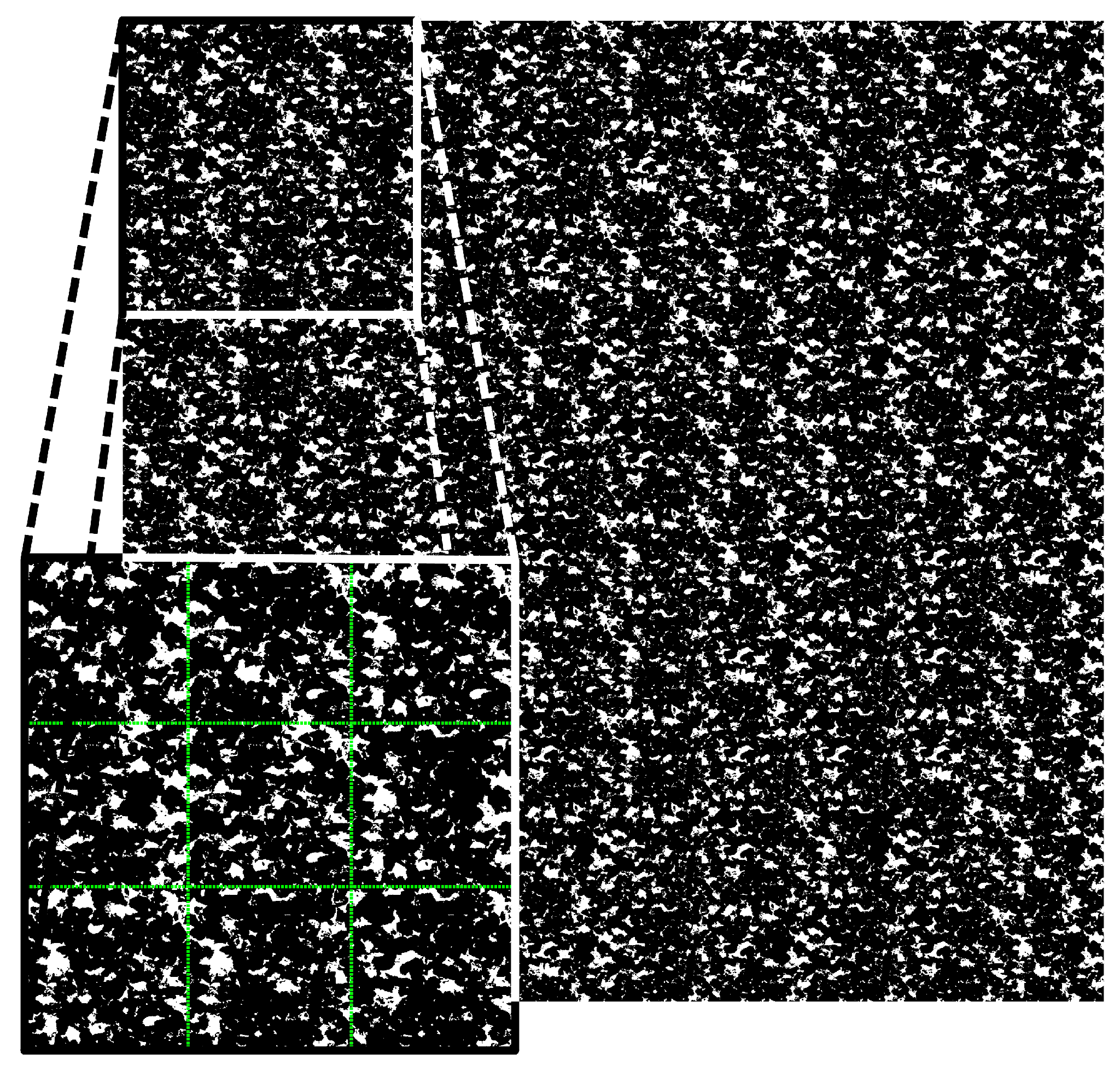}&
    \includegraphics[width=0.48\columnwidth]{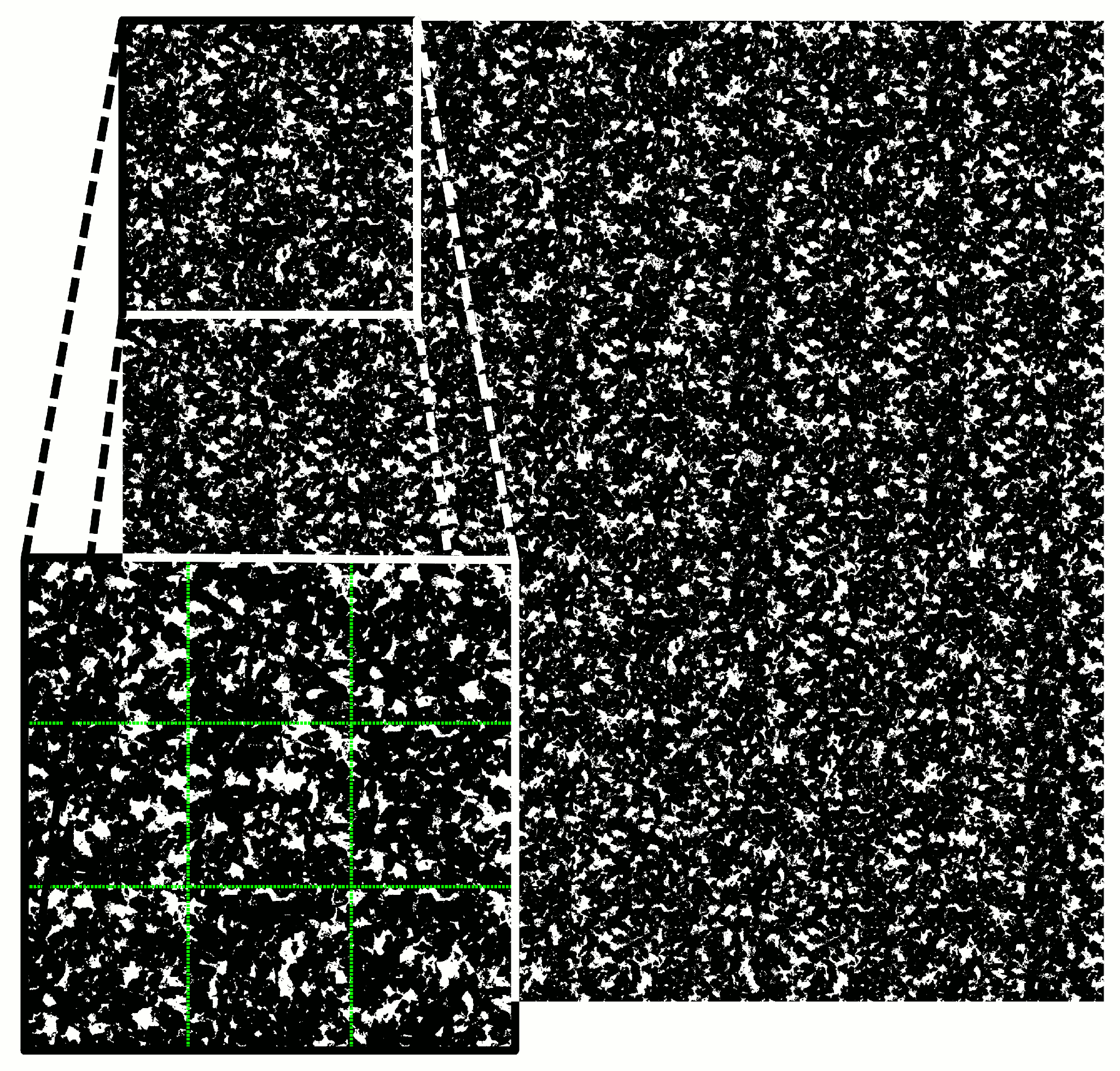}\\
    (a) & (b)
  \end{tabular}
  \caption{Example of synthesized \sandstone{} microstructure of $3\times3$ unpatched (a) and patched (b) tiles.}
  \label{fig:sinth_microstructures_patched}
\end{figure}

\begin{figure}[htb!]
  \centering
  \setlength{\tabcolsep}{0.00cm}
  \begin{tabular}{cc}
    \includegraphics[width=0.50\columnwidth]{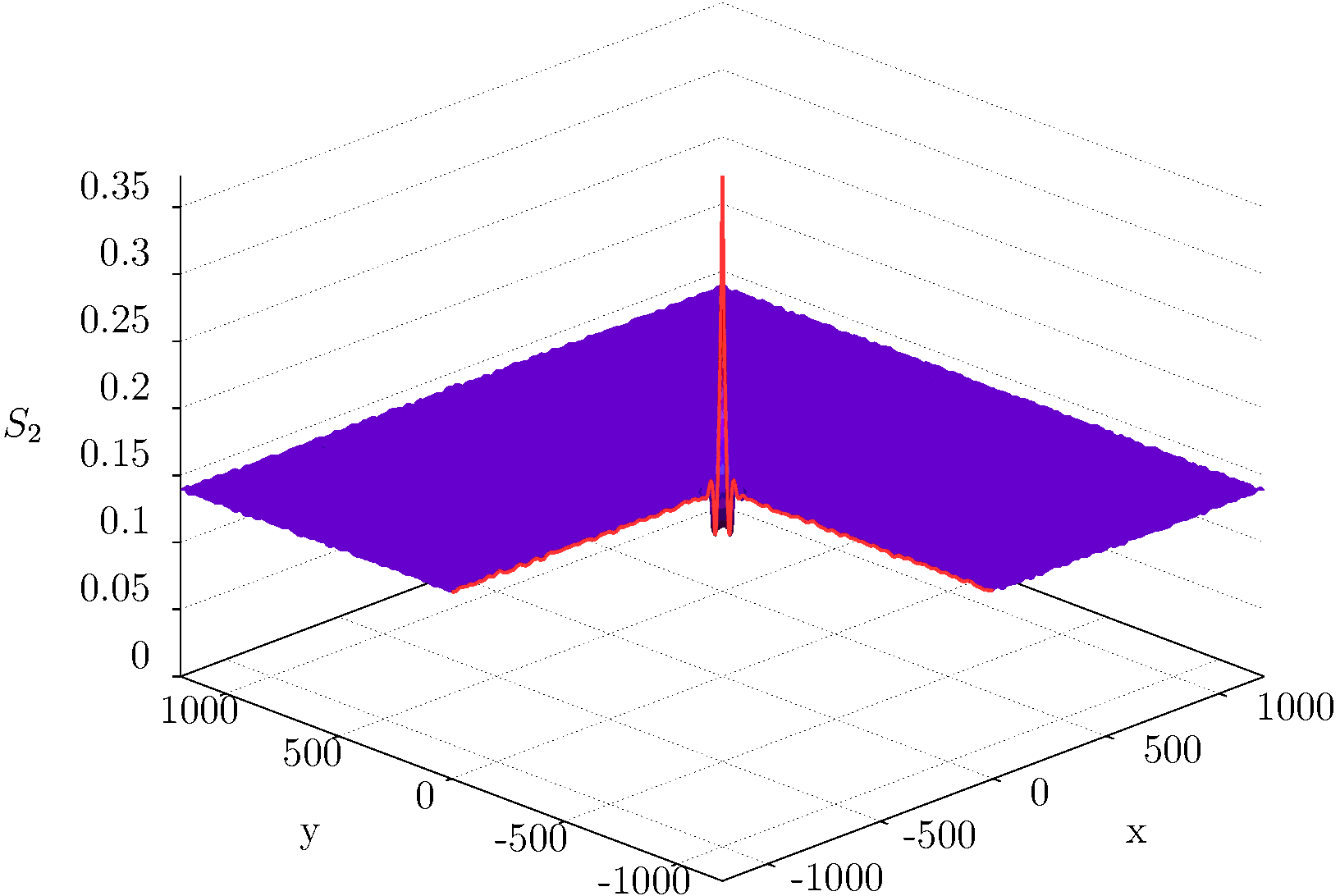} &
    \includegraphics[width=0.50\columnwidth]{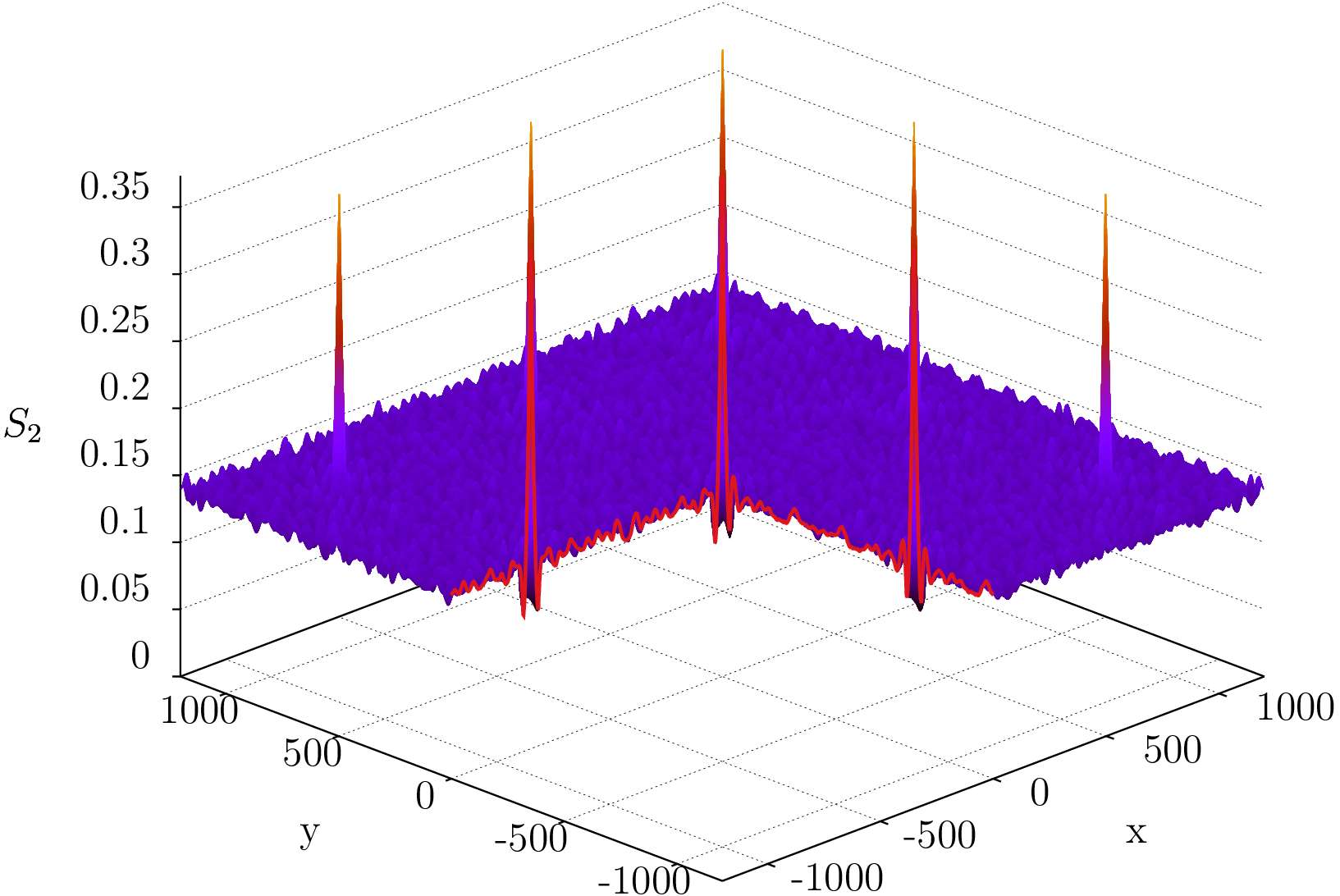}\\
    (a) &(b)\\
    \includegraphics[width=0.50\columnwidth]{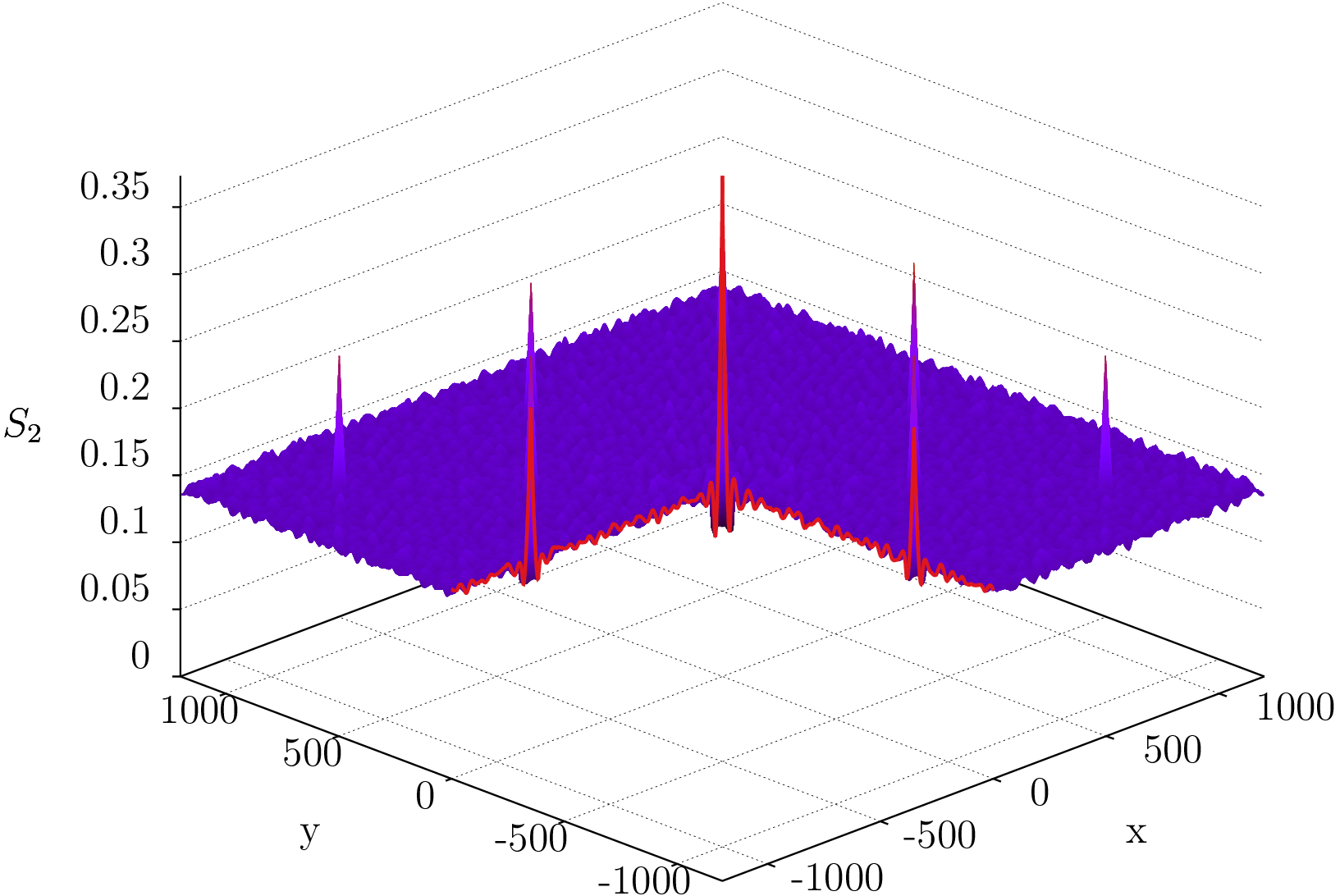}&
    \includegraphics[width=0.50\columnwidth]{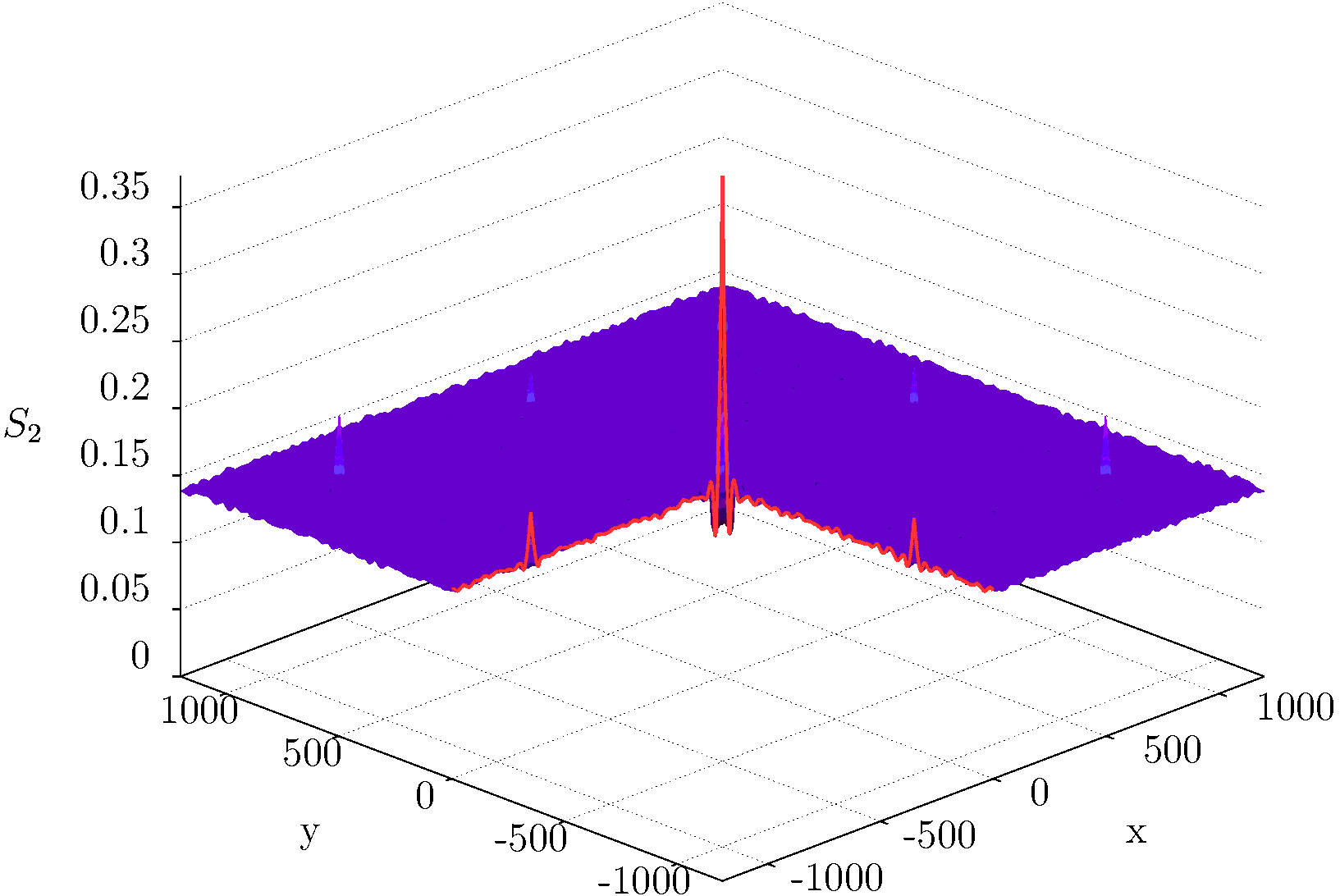}\\
    (c) &(d)
  \end{tabular}
  \caption{(Colour online) \SII{} statistics of \hdisks{} microstructure, a) target system, b) \Wa{1}{1}{1} (\puc{}), c) \Wa{16}{2}{2}, d) \Wa{16}{2}{2} with patches}
  \label{fig:S2comparisson}
\end{figure}

Following the above indicators, one can select the optimal set with respect to the desired compression capabilities and the level of induced degeneracy in terms of spurious long range orientation orders quantified by $\Se$. For instance, examples of synthesized microstructures of target systems from~\Fref{fig:target_systems} are displayed in~\Fref{fig:sinth_microstructures}. These particular reconstructions are formed by tilings made up of $10\times 10$ tiles. The zoomed left upper corners contain only $3\times 3$ tiles for better visualization of short range features.
In \Fref{fig:sinth_microstructures_patched}, we further show synthesized \sandstone{} microstructures created by means of unpatched and patched tiles of the set \Wa{18}{3}{3} displayed at the two top rows.
The conjectures coming from~\Fref{fig:sets_vs_S2}b are very difficult to follow by visual inspection, however the patched reconstruction in \Fref{fig:sinth_microstructures_patched}b seems to us less polluted by repetitive patterns then that without the patches, see also Figs.~\ref{fig:sinth_microstructures_patched}a and \ref{fig:S2comparisson}.

\section{Conclusions and future developments}\label{s:conclusions}

In general, the microstructure compression is subject to a compromise among various factors. The design of a compression technique calls into question the degree of compression (dimensions of tiles and tile set cardinality), the amount of distortion induced (parasitic long range orientation orders along with the short range defects in compressed microstructures, e.g. due to quilting), and the computational overhead required to compress and uncompress the data.

\begin{figure}[htb!]
   \centering
   \begin{tabular}{cc}
     \includegraphics[width=0.60\columnwidth]{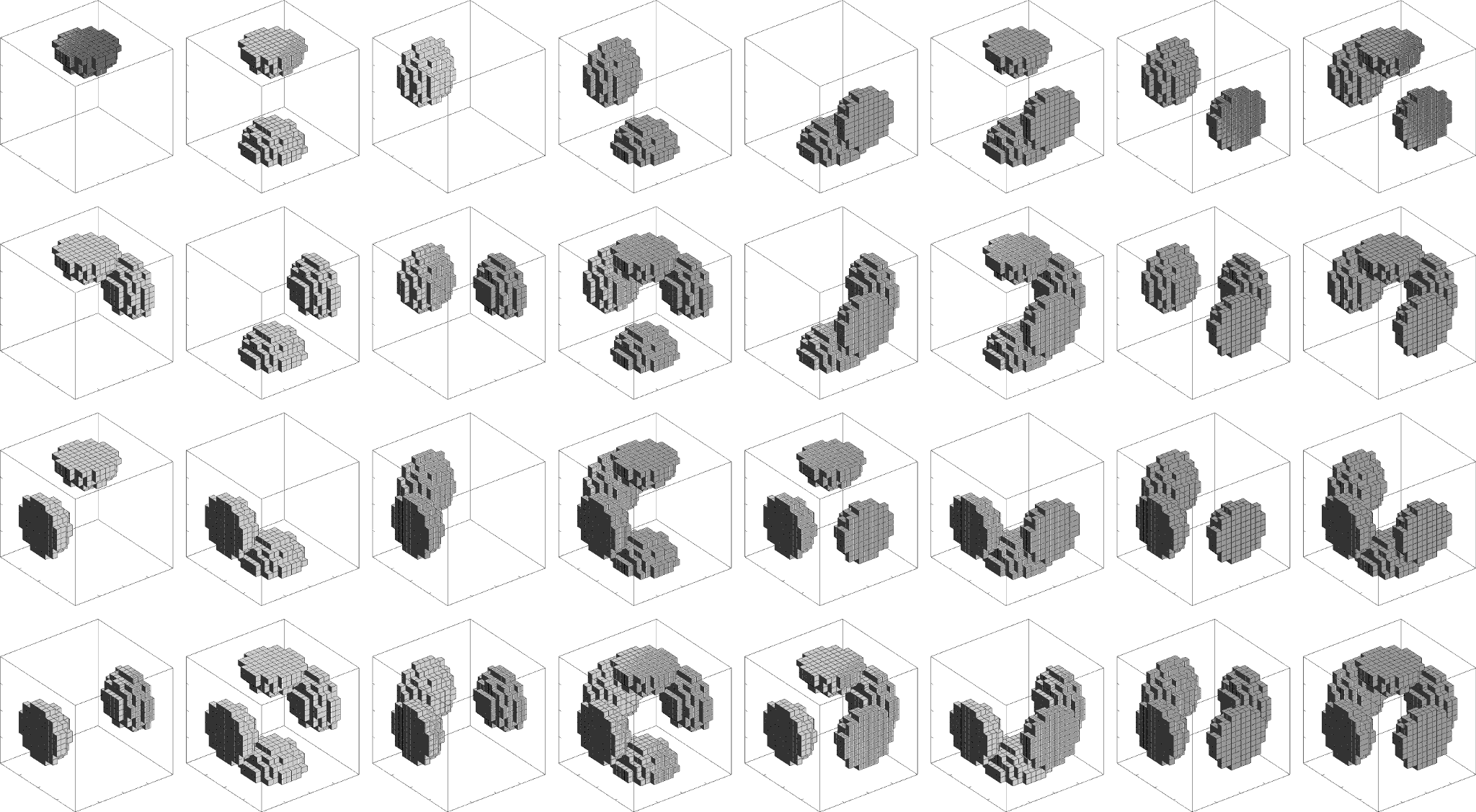}&
     \includegraphics[width=0.35\columnwidth]{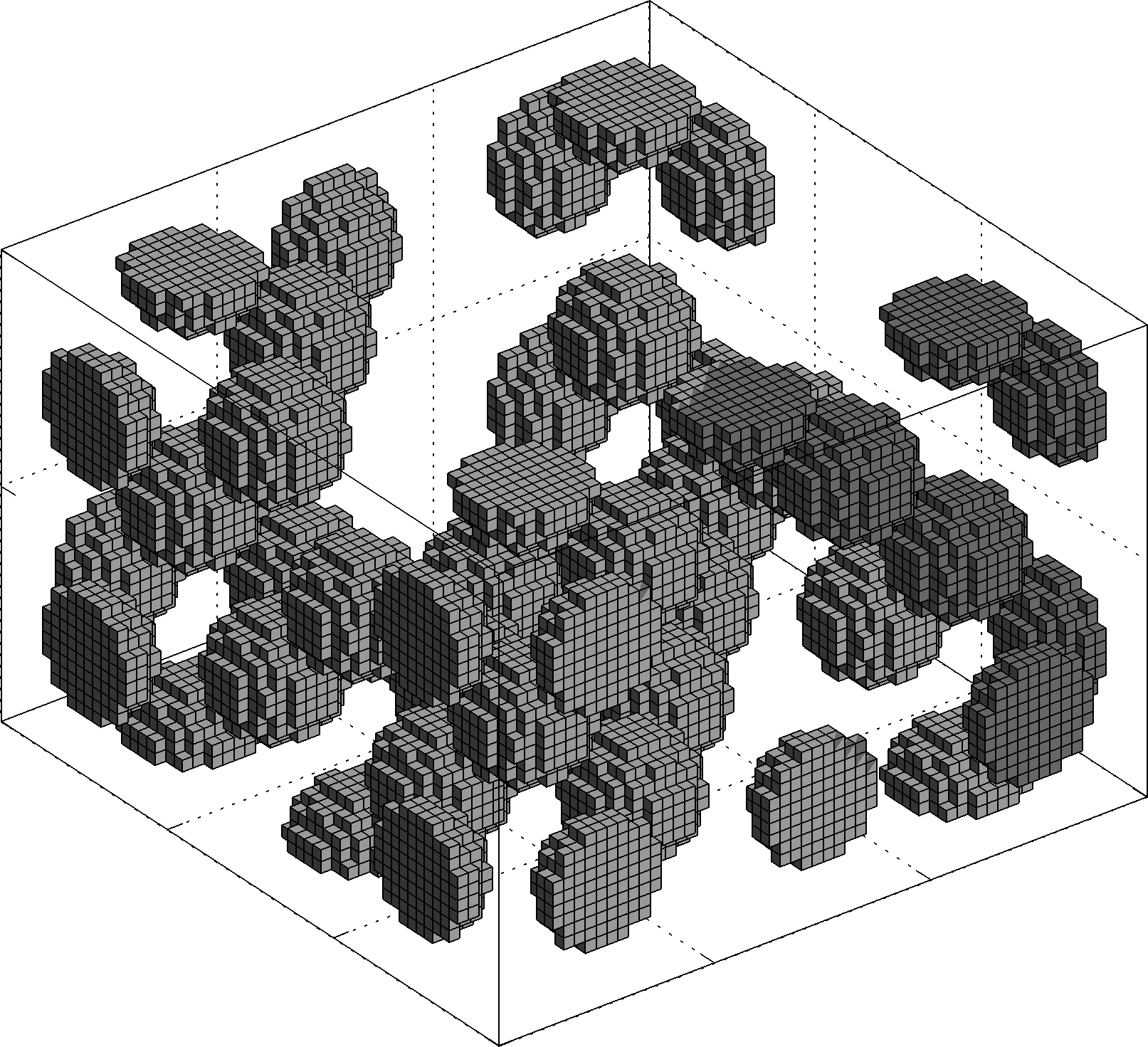}\\
     (a) & (b)\\
   \end{tabular}
   \caption{a) tile set W3/3-3-3, b) tiling consisting of $3\times3\times2$ tiles}
   \label{fig:3D_tiling}
 \end{figure}

In this work, we have proposed an approach to compression and reverse synthesis of microstructural patterns of real world microstructures based on Wang tilings, image processing techniques, and statistical quantification. The method goes beyond periodic representations, and provides a natural generalization of \puc{} concept (recall, \Wa{1}{1}{1} equals \puc{}). It allows to represent complex microstructural patterns by making use of small data sets called the Wang tile sets. The properties of the automatic tile morphology design were investigated by means of a number of sensitivity analyses whose objective was to determine the optimal values of the input parameters, such that the compressed microstructure contains maximum microstructural information and is small enough for an inexpensive treatment. From our results, we conjecture that the width of the overlapping region of about nine times the mean characteristic inclusion size was suitable for all investigated microstructures. However, no general rule regarding the input values was observed besides. A similar sensitivity study, following the flowchart in~\Fref{fig:param_flowchart}, is therefore recommended any time the compression based on Wang tiles is desired. The extension of the concept to three dimensional setting by means of Wang cubes is fairly straightforward and is in the focus of our future work. Preliminary outcomes are demonstrated by an example shown in ~\Fref{fig:3D_tiling}.

\paragraph{Acknowledgements}

The authors acknowledge the Czech Science Foundation, Grant No. 13-24027S and the European Social Fund endowment under Grant No. CZ.1.07/2.3.00/30.0005 of Brno University of Technology (Support for the creation of excellent interdisciplinary research teams at Brno University of Technology).

\end{document}